\documentclass[a4paper,10pt]{article}
\pdfoutput=1 
\usepackage{t1enc}
\usepackage[utf8x]{inputenc}
\usepackage[english]{babel}
\usepackage{pdfrack}
\usepackage{psfrag}
\usepackage{amsmath}
\usepackage{amsmath}
\usepackage{anysize}
\usepackage{amssymb}
\usepackage{stmaryrd}
\usepackage[amsthm]{ntheorem-hyper}
\usepackage{subfloat}
\usepackage{subfig}
\usepackage{caption}
\DeclareMathOperator{\I}{I}
 \DeclareMathOperator{\hH}{H}

\def\iX{{\cal X}}
\def\iY{{\cal Y}}
\newcommand{\eps}{\varepsilon}
\newcommand{\Prob}{\textnormal{Pr}}
\newcommand{\markov}{\minuso}
\newcommand{\irott}{\mathcal}
\newcommand{\Vx}{\mathbf x}
\newcommand{\vX}{\mathbf X}
\newcommand{\vY}{\mathbf Y}

\newcommand{\VD}{\mathbf D}
\theoremstyle{plain}
\newtheorem{Thm}{Theorem}

\newtheorem{Cor}{Corollary}
\theoremstyle{remark}
\newtheorem{Rem}{Remark}
\theoremstyle{plain}
\newtheorem{lemma}{Lemma}
\theoremstyle{definition}
\newtheorem{Def}{Definition}
\newtheorem{Exa}{Example}
\theoremstyle{break}

\marginsize{15mm}{15mm}{7mm}{7mm}

\title{On Capacity Regions of Discrete Asynchronous Multiple Access Channels }
\author{L. Farkas,
        T. Kói
\thanks{It has been presented in part at ISIT 2011, Saint Petersburg. L\'or\'ant Farkas is with the Department
of Analysis, Budapest University of Technology and Economics,
e-mail: lfarkas@math.bme.hu. Tamás Kói is with the Department
of Stochastics, Budapest University of Technology and Economics and with the MTA-BME Stochastics Research Group,
e-mail: koitomi@math.bme.hu.}}

\begin{document}

\maketitle

\begin{abstract}
A general formalization is given for asynchronous multiple access channels which admits different assumptions on delays. This general framework allows the analysis of so far unexplored models leading to new interesting capacity regions. The main technical result is the single letter characterization for the capacity region in case of 3 senders, 2 synchronous with each other and the third not synchronous with them.
\vspace{3mm}
\vspace{3mm}
\end{abstract}

\section{Introduction}
Ahlswede \cite{Ahlswede} and Liao \cite{liao}  showed that if two senders communicate synchronously over a discrete memoryless multiple access channel (MAC) which is characterized by a stochastic matrix $W(y|x_1,x_2)$, it is possible to communicate with arbitrary small average probability of error if the rate pair is inside the following pentagon:
  \begin{align}
   0\leq R_1&\leq I(X_1 \wedge Y|X_2) \notag\\
   0\leq R_2&\leq I(X_2 \wedge Y|X_1) \notag\\
   R_1+R_2&\leq I(X_1,X_2 \wedge Y) \label{pentagon1}
  \end{align}
for some independent input random variables $X_1$, $X_2$, where $P(Y=y|X_1=x_1,X_2=x_2)=W(y|x_1,x_2)$. Moreover, the convex hull of the union of these pentagons can also be achieved, via time sharing, while no rate pair outside this convex hull is achievable.

The discrete memoryless asynchronous multiple access channel (AMAC) arises when the senders can not synchronize the starting times of their codewords, rather, there is an unknown delay between these starting times. Cover, McEliece and Posner \cite{Cover} showed that if the delay is bounded by $b_n$ depending on the codeword length $n$ such that $\frac{b_n}{n}\to 0$ then the convex closure is still achievable by a generalized time sharing method.

Poltyrev \cite{Poltyrev} and Hui and Humblet \cite{Hui-Humblet} addressed models with arbitrary delays known (in \cite{Poltyrev}) or unknown (in \cite{Hui-Humblet}) to the receiver. For such models, the capacity region was shown to be the union of the pentagons above although with some gaps in the proofs, see Appendix A. Verdú \cite{Verdu} studied asynchronous channels with memory. His model slightly differs from common models: the time runs over a torus rather than from $-\infty$ to $\infty$.  Later, Grant, Rimoldi, Urbanke and Whiting in \cite{Urbanke} showed that in the informed receiver case the union can be achieved by rate splitting and successive decoding. The gap in the achievability proof of \cite{Hui-Humblet} for the uninformed receiver case has been filled in the book of El Gamal and Kim \cite{elgamal}.

This paper is an extended version of the ISIT 2011 contribution \cite{isit2011}, originating from the authors' effort to derive the AMAC capacity region without gaps in the proof (the result in \cite{elgamal} was unknown to us at the time, as was, apparently, to the reviewers of \cite{isit2011}). More than doing that, in \cite{isit2011} a general formalization for AMACs was introduced, allowing dependence of the capacity region on the distribution of the delays, typically through the support of that distribution. For a particular (somewhat artificial) choice of the delay distribution, the capacity region was determined, providing the first example that the capacity region could be strictly between the union and its convex closure.

The main technical result of this paper is new compared to \cite{isit2011}. It is a single letter characterization of the capacity region for 3 senders, two synchronized with each other and the third unsynchronized with them.

In section 2 we give the formal description of the AMAC model, where several possible definitions are given, which are analyzed in parallel. In Section 3 a general converse is presented. This converse is used when capacity region of known (in section 4) and previously unknown (in sections 5 and 6) models are derived.

Our achievability proofs rely on the techniques of rate splitting and successive decoding developed by Grant, Rimoldi, Urbanke, Whiting \cite{Urbanke} and Rimoldi \cite{rimoldi}.

\section{Model of coding for the AMAC}\label{Bevezetés}
In this paper vectors (finite sequences) will be denoted by boldface symbols. Furthermore, $\left[ i \right]$ denotes $\left\{ 1,2, \dots, i \right\}$.

A $K$-senders asynchronous discrete memoryless multiple-access channel ($K$-AMAC) is defined in terms of $K$ finite input alphabets $\mathcal{X}_i, i \in \left[ K \right]$, a finite output alphabet $\mathcal{Y}$, and a stochastic matrix $W: \mathcal{X}_1 \times \mathcal{X}_2 \times \dots \times  \mathcal{X}_{K} \rightarrow \mathcal{Y}$ describing the probability distribution of the output given the inputs.

\begin{Def}
A codebook system of block-length $n$ with rate vector $\mathbf{R}=(R_1,R_2 \dots, R_K)$ for a given $K$-AMAC $W$ consists of $K$ codebooks $C_1, C_2, \dots, C_K$, where the codebook $C_m$ of the $m$-th sender has $2^{nR_m}$ codewords of length $n$ whose symbols are from $\mathcal X_m$.
\end{Def}

The system is symbol synchronized but not frame synchronized. The differences between the timing of the receiver and the timings of the senders are represented by a K-tuple of delays as in Definition \ref{delay}.

The senders have two-way infinite sequences of random messages, and assign codewords to their consecutive messages. The codewords go through the channel. The sequences of the senders' codewords and hence also the output symbol sequence are two-way infinite sequences. Fix the location of the  0-th output symbol. The message of sender $m \in \left[ K \right]$ whose codeword affects the $0$'th output is denoted by $M_{m,0}$. This restricts the delays to be in the set $\{0,1,\dots,n-1\}$. Formally, we  use the following definitions:
\begin{Def} \label{kodszimbolum}
For each integer $j \in \mathbb Z$ and for each  $m \in \left[ K \right]$ let $M_{m,j}$ be a uniformly distributed random variable taking values in the set $\{1,2, \dots, 2^{nR_m} \}$. All these random variables are independent of each other. The two-way infinite sequence $\{ M_{m,j}, j \in \mathbb Z \}$ represents the message flow sent by the $m$-th sender. For each integer $j \in \mathbb Z$ and for each $m \in \left[ K \right]$ let $\mathbf X_{m,j}$ denote the $M_{m,j}$-th codeword in the codebook of sender $m$. Let $X_{m,nj+i}$ be the $i$-th symbol of $\mathbf X_{m,j}$ where $i \in \{ 0,1, \dots, n-1 \}$.
\end{Def}
\begin{Def} \label{delay}
For each $n \in \mathbb Z^{+}$, let $$\mathbf{D}(n) = \left( D_1(n),D_2(n),\dots,D_K(n) \right)$$ be a K-tuple of random variables, not necessarily independent of each other but independent of all previously defined random variables, taking values in the set $\{0,1,\dots,n-1\}$. $D_m (n)$ will represent the delay of sender $m$ relative to the receiver's timing. The joint distribution of delays is known to the senders and the receiver. The realizations of the random variables $D_1(n),D_2(n),\dots,D_K(n)$ are not known to the senders and, depending on the model, may be known or unknown to the receiver. The sequence $\mathbf{D}= \{ \mathbf{D}(1),\mathbf{D}(2), \dots, \mathbf{D}(n), \dots \}$ will be called the delay system. With a slight abuse of notation, we also write $\mathbf{D}$ instead of $\mathbf{D}(n)$.
\end{Def}
\begin{Rem}
Our definition allows arbitrary distributions for the delays for each blocklength $n$. Clearly, in practical models these distributions can not be arbitrary, but have to satisfy consistence conditions. We have chosen this general model since we think that any practical model can be described this way.
\end{Rem}
\begin{Exa} \label{pelda1}
For each $n \in \mathbb Z^{+}$ and for each $m \in \left[ K \right]$ $D_m (n)$ has uniform distribution on $\{0,1,\dots,n-1\}$ and they are independent. Following \cite{Hui-Humblet} it is called the totally asynchronous case in the paper.
\end{Exa}
\begin{Exa} \label{pelda2}
Let $K=2$, for each $n \in \mathbb Z^{+}$ let $D_1 (n)$, $D_2 (n)$ be independent random variables uniformly distributed on the even numbers of $\{0,1,\dots,n-1\}$. It is called the even delays case in the paper.
\end{Exa}
\begin{Exa} \label{pelda3}
Let $K=3$, for each $n \in \mathbb Z^{+}$ let $D_1(n)=D_2(n)$ be a random variable uniformly distributed on $\{0, 1, \dots, n-1\}$ and let $D_3(n)$ be a random variable independent of $D_1(n)$ and uniformly distributed on $\{0, 1, \dots, n-1\}$. It is called the partly asynchronous three senders case in the paper.
\end{Exa}

For fixed $n$, the output sequence is defined as follows:
\begin{Def} \label{kimenet}
Let $Y_{nj+i}$ be the output random variable of the channel with transition matrix $W$ when the inputs are $X_{1,nj+i+D_1(n)}$, $X_{2,nj+i+D_2(n)}$, $\dots$, $X_{K,nj+i+D_K(n)}$ where $i \in \{ 0,1, \dots, n-1 \}$.
\end{Def}

It is possible to define the decoder in several ways. We will consider two different definitions, which give the strongest version of the converse and direct parts of the coding theorems, respectively.

\begin{Def} \label{dekodol2}
An informed infinite decoder is defined as a function which assigns to each two way infinite output sequence realization $\{ y_l, l \in \mathbb Z \}$ and each realization of $ \mathbf{D}(n) = \left( D_1(n),D_2(n),\dots,D_K(n) \right), n \in \mathbb Z^{+} $, a K-tuple of messages $\{ \hat m_{m,0}, m \in \left[ K \right] \}$.
\end{Def}
\begin{Def} \label{dekodol3}
An uninformed L-block decoder, $L \in \mathbb Z^{+}$, is defined as a function which assigns to each $(2Ln+1)$-tuple $\{ y_{l}, l \in \{-Ln, \dots, 0, \dots, Ln \} \}$ of possible output realizations a K-tuple of messages $\{ \hat m_{m,0}, m \in \left[ K \right] \}$.
\end{Def}

The codebooks and the decoder form an $n$-length coding/decoding system.

\begin{figure}[Hbt]
\begin{center}
\includegraphics[width=8.5cm]{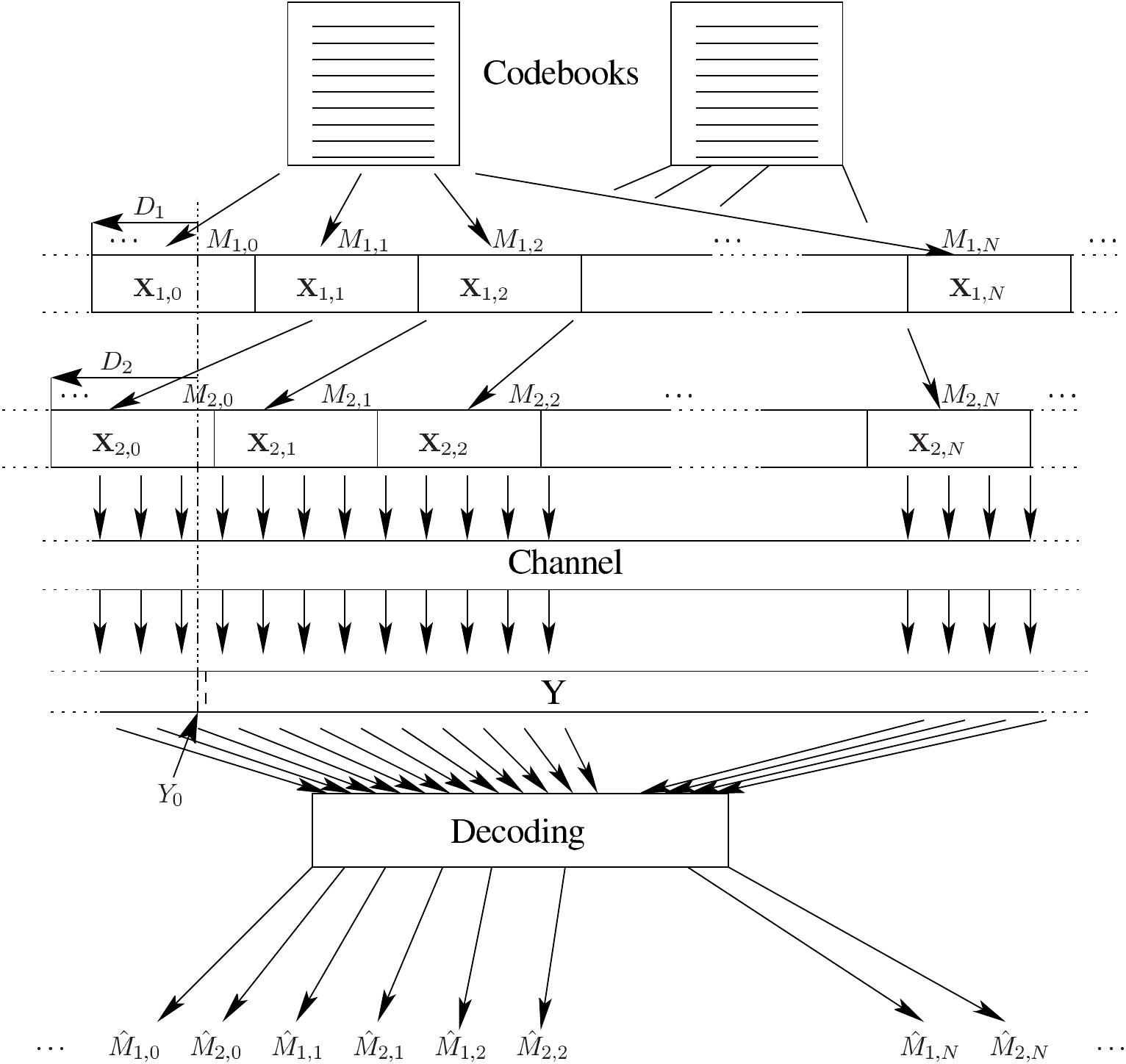}
\caption{ The Setting for Two Senders }
\label{fig-0}
\end{center}
\end{figure}

The definitions above determine the probability structure of the model. For each $m$ the random variable sequence $\{ M_{m,j}, j \in \mathbb Z \}$ is the two way infinite message flow of the $m$-th sender. The corresponding flow of codewords is  $\{ \mathbf X_{m,j}, j \in \mathbb Z \}$. The flows of the senders, the channel transition and the delay system $\mathbf{D}$, define a two way infinite output random variable sequence $\{ Y_l, l \in \mathbb Z \}$. In case of uninformed $L$-block decoder the receiver examines the output block $Y_{-Ln},Y_{-Ln+1}, \dots, Y_0,Y_1, \dots Y_{Ln}$ from which estimations $\{ \hat M_{m,0}, m \in \left[ K \right]$ are created. In case of an informed infinite decoder the whole output sequence and the realizations of delays are used in the estimations $\{ \hat M_{m,0}, m \in \left[ K \right] \}$. It is assumed that the same but shifted decoding procedure occurs at the output points $\{ nk, k \in \mathbb Z \}$. Hence the random variables of the estimations $\{ \hat M_{m,j}, m \in \left[ K \right],j \in \mathbb Z \}$ are also defined. See Fig. \ref{fig-0}. for this model, in case $K=2$.

We will consider two different error definitions. As standard for multiple-access channels, both errors are average error over messages. However, our first error type also involves averaging over delays, while the second one takes maximum over the possible delays. In this paper, the terms average error and maximal error will be used as defined below. 
\begin{Def}
The average error is the following:
\begin{equation} \label{hibatag}
P^n_e=\textnormal{Pr}\left\{ \bigcup_{m=1}^K \left\{ M_{m,0}\neq \hat M_{m,0} \right\} \right\}.
\end{equation}
\end{Def}
\begin{Def} \label{maximal}
The maximal error is the following:
\begin{equation}
P^n_e (*)= \max_{\mathbf{d}(n): \textnormal{Pr}\left\{  \mathbf{D}(n) = \mathbf{d}(n)  \right\} > 0} \textnormal{Pr}\left\{ \bigcup_{m=1}^K \left\{ M_{m,0}\neq \hat M_{m,0} \right\} | \mathbf{D}(n)=\mathbf{d}(n) \right\}.
\end{equation}
\end{Def}
\begin{Rem}
The average error depends on the joint distribution of delays  $\left( D_1(n),D_2(n),\dots,D_K(n) \right)$, while the maximal error depends on the joint distribution of the delays only through its support.
\end{Rem}
\begin{Rem}
The two kinds of error are related very closely. If $P^n_e (*) \rightarrow 0$ then $P^n_e \rightarrow 0$. On the other hand, if $P^n_e \rightarrow 0$ exponentially  as $n \rightarrow \infty$ and if $\min_{\mathbf{d}(n): \textnormal{Pr}\left\{  \mathbf{D}(n) = \mathbf{d}(n)  \right\} > 0} \textnormal{Pr}\left\{  \mathbf{D}(n) = \mathbf{d}(n)  \right\}$ tends to $0$ slower than exponentially then also $P^n_e (*) \rightarrow 0$ exponentially.
\end{Rem}

We have defined several types of models according to the various definitions of decoder and of error. For the sake of brevity, the following definition is meant to define a capacity region simultaneously for all cases. Here, in case of $L$-block decoder, a proper choice of $L$ is understood. In particular cases, a suitable $L$ will be specified, not entering the question whether a smaller $L$ would also do.
\begin{Def} \label{kapacitas}
Corresponding to the delay system $\mathbf{D}$, the rate vector $(R_1,R_2,\dots,R_K)$ is achievable if for every $\eps >0$, $\delta >0$  for all $N \in \mathbb Z^{+}$ there exists a coding/decoding system with blocklength $n>N$ with rates coordinate-wise exceeding $(R_1 - \delta,R_2 -\delta,\dots,R_K - \delta)$ and with error less than $\eps$. The set of achievable rate vectors is the capacity region of the $K$-AMAC.
\end{Def}
\begin{Rem}
In the definition above we used the 'optimistic' definition of capacity region, rather than the more usual 'pessimistic one', see \cite{Csiszar}\footnote{In short, in the 'optimistic' definition it is enough to show that there is a "good" coding/decoding system for a sequence of blocklength $n_k \to \infty$.}. The reason is that in the even delays case there are differences in the performance of coding/decoding systems of even and odd blocklength (see Theorem \ref{parosconverse}).
\end{Rem}
\begin{Rem}
If for some region achievability is proved in case of uninformed $L$-block decoder with maximal error, and the converse is proved in case of informed infinite decoder with average error, then for any combination of the model assumptions above the capacity region is equal to this region.
\end{Rem}

\begin{lemma} \label{alarak}
For either type of AMAC model, if $\mathbf{D}$ and $\mathbf{D}'$ are two delay systems such that for some $0 < \alpha \le 1$ for all $n \in \mathbb Z^{+}$ and $\mathbf{d}(n) \in \{0,1,\dots,n-1 \}^{K}$
\begin{equation} \label{alarakas}
\textnormal{Pr}\left\{  \mathbf{D'}(n) = \mathbf{d}(n)  \right\} \ge \alpha \textnormal{Pr}\left\{  \mathbf{D}(n) = \mathbf{d}(n)  \right\},
\end{equation}
then the capacity region under delay system $\mathbf{D}'$ is contained (perhaps strictly) in the capacity region under delay system $\mathbf{D}$.
\end{lemma}
\emph{Proof:}
Consider an arbitrary $n$ length coding/decoding system. Then $P^n_{e,\mathbf{D'}(n)} \ge \alpha P^n_{e,\mathbf{D}(n)}$ and $P^n_{e,\mathbf{D'}(n)} (*) \ge P^n_{e,\mathbf{D}(n)} (*)$ hold, where the lower indices indicate the underlying delay system. This proves the lemma. $\blacksquare$

\begin{Rem}
In case of any type of decoder, if two delay systems $\mathbf{D}$ and $\mathbf{D}'$ have the same support set for each $n$, then the capacity regions corresponding to delay systems $\mathbf{D}$ and $\mathbf{D}'$  coincide in case of maximal error. Furthermore, if the equation (\ref{alarakas}) is fulfilled by $\mathbf{D}$ and $\mathbf{D}'$ and it is also fulfilled when the roles of $\mathbf{D}$ and $\mathbf{D}'$ are reversed, then by Lemma \ref{alarak} the capacity regions also coincide in case of average error.
\end{Rem}

\section{A general converse}\label{Converse}
In this section a general converse theorem is proved, which depends on the delay system. In the following sections, this general converse is used to derive the capacity region of special cases.

For all subset $S$ of $\left[ K \right]$ write
\begin{equation}
\mathbf{X}_{S}=(X_i)_{i \in S} \text{, }S^{c} = \left[ K \right] \setminus S,
\end{equation}
and for all $\mathbf{R}=(R_1,R_2, \dots, R_K)$ write
\begin{equation}
R(S)= \sum_{i \in S} R_{i}.
\end{equation}

Let $\mathbf D$ denote the delay vector. Let $\mathbf X_{B,i+D_B}$ denote the random vector with components $X_{l,i+D_{l}}$, $l\in B$ where $B$ $\subset$ $[K]$ and $X_{m,j}$ is defined as in definition \ref{kodszimbolum}; similar notation is used where $+$ is replaced by $\oplus$ which means addition modulo $n$.
\begin{Thm}\label{main-converse-prop-alt} For any $n$ length coding/decoding system for a $K$ senders AMAC $W$ with informed infinite decoder, the following bounds hold for the rate vector $\mathbf{R}=(R_1,R_2,  \dots, R_k)$ for all $S \subset \left[ K \right]$:
\begin{equation} \label{upperbound}
R(S) \leq \I(\mathbf X_{S,Q\oplus D_S} \wedge \tilde Y_Q|\mathbf X_{S^c,Q \oplus D_{S^c}},Q,\VD) +  \eps_n.
\end{equation}
Here $\eps_n= (R(\left[ K \right]))P_e^{n}+\frac{1}{n}$, the random variable $Q$ is uniformly distributed on $\{0,1,\dots,n-1\}$ and independent of $\mathbf D$ and the message flows of the senders. Further, $\tilde Y_Q$ is linked to the random variables $X_{1,Q\oplus D_1},X_{2,Q \oplus D_2}, \dots, X_{K,Q \oplus D_K}$ through the channel $W$; formally, its conditional distribution given $Q,\mathbf{D}$ and $X_{1,Q\oplus D_1},X_{2,Q \oplus D_2}, \dots, X_{K,Q \oplus D_K}$ depends only on the values $x_1,x_2,\dots, x_K$ of the latter random variables and is equal to $W(\cdot|x_1,x_2,\dots,x_K)$.
\end{Thm}
\begin{Rem}
Theorem \ref{main-converse-prop-alt} will be used for sequences of coding/decoding systems with $P_e^{n} \rightarrow 0$. In this case $\eps_n$ also tends to $0$.
\end{Rem}
\emph{Proof:}
For the sake of clarity just the two senders special case is addressed here, the full proof of Theorem \ref{main-converse-prop-alt} can be found in Appendix B. In case of two senders the bounds (\ref{upperbound}) are:
\begin{align} \label{felsobecsles1}
R_1&\leq \I(X_{1,Q\oplus D_1} \wedge \tilde Y_{Q}|X_{2,Q\oplus D_2},Q,D_1,D_2)+\eps_n\\ \label{felsobecsles2}
R_2&\leq \I(X_{2,Q\oplus D_2} \wedge \tilde Y_{Q}|X_{1,Q\oplus D_1},Q,D_1,D_2)+\eps_n\\ \label{felsobecsles3}
R_1+R_2 &\leq \I(X_{1,Q\oplus D_1},X_{2,Q\oplus D_2}\wedge \tilde Y_{Q}|Q,D_1,D_2)+\eps_n
\end{align}
Note that $n\eps_n=n(R_1+R_2)P_e^{(n)}+1$. Hence
\begin{equation}
 n\eps_n \geq \hH(M_{1,0},M_{2,0}|\hat M_{1,0},\hat M_{2,0}) \label{Fano}
\end{equation}
by Fano's inequality.

\begin{figure}[t]
 \begin{center}
 \includegraphics[width=8.5cm]{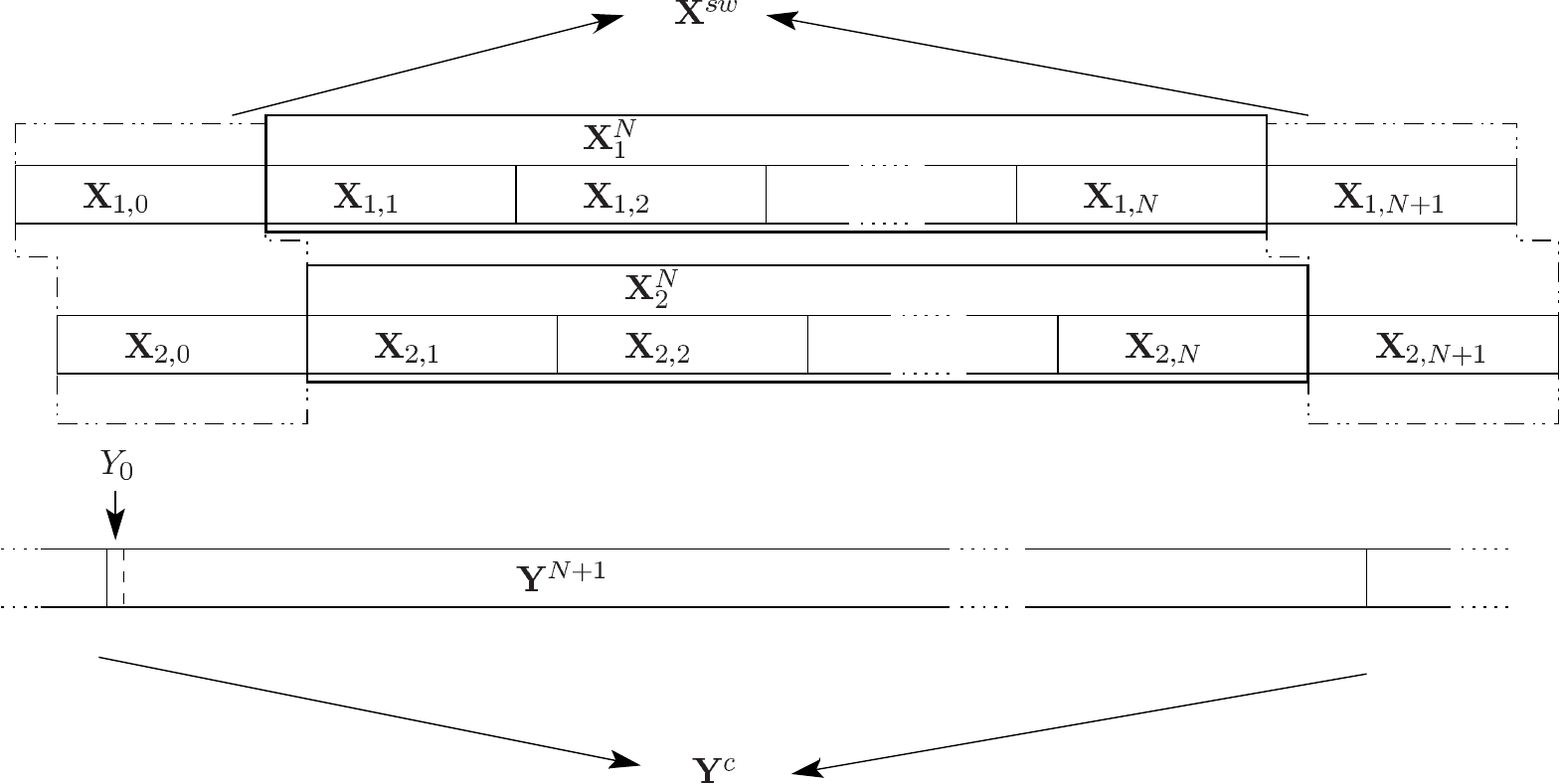}
 \caption{ The random variables that play role in the bound on the sum $R_1+R_2$}
 \label{fig-2}
 \end{center}
\end{figure}

We just bound $R_1+R_2$. The bounds for $R_1$ and $R_2$ can be derived similarly (See also Appendix B).

Take a window of the receiver consisting of $N+1$ $n$-length blocks $\mathbf Y^{N+1}=\{ Y_0,Y_1,\dots Y_{n(N+1)-1} \}$. This window fully covers the code-blocks $\vX_{1,1}$,$\vX_{1,2},\dots$,$\vX_{1,N}$ of sender 1 and $\vX_{2,1}$,$\vX_{2,2},\dots$,$\vX_{2,N}$ of sender 2, denoted by $\mathbf X_1^{N}$ and $\mathbf X_2^{N}$ respectively. The codewords at the sides of the output window are $\vX_{1,0},\vX_{2,0},\vX_{1,N+1},\vX_{2,N+1}$, denote this quadruple by $\vX^{sw}$ (the expression "side of the windows" is abbreviated by the index $sw$). Then we have

\begin{align}
Nn &(R_1+R_2)=\label{2} \hH(\mathbf M_1^{N},\mathbf M_2^{N})\\
 =&\I(\mathbf M_1^{N},\mathbf M_2^{N} \wedge \hat{\mathbf{M}}_1^{N},\hat{\mathbf{M}}_2^{N}) +\hH(\mathbf M_1^{N},\mathbf M_2^{N}|\hat{\mathbf{M}}_1^{N},\hat{\mathbf{M}}_2^{N}) \\
 \leq& \I(\mathbf M_1^{N},\mathbf M_2^{N} \wedge \hat{\mathbf{M}}_1^{N},\hat{\mathbf{M}}_2^{N}) +\sum_{i=1}^{N} \hH(M_{1,i},M_{2,i}|\hat M_{1,i},M_{2,i})\\
 \leq& \I(\mathbf M_1^N,\mathbf M_2^{N} \wedge \hat{\mathbf{M}}_1^N,\hat{\mathbf{M}}_2^{N}) + Nn\eps_n \label{2a}\\
 \leq& \I(\vX_1^{N},\vX_2^{N} \wedge \vY^{N+1},\vX^{sw},D_1,D_2 ) + Nn\eps_n \label{2b}
\end{align}
where (\ref{2a}) comes from (\ref{Fano}) and (\ref{2b}) comes from the Markov relation
\begin{align}
&(\mathbf M_1^N,\mathbf M_2^{N})\markov (\mathbf X_1^N,\mathbf X_2^{N})\markov (\mathbf Y^{N+1},\vX^{sw},D_1,D_2) \markov\notag\\
&\markov (\vY^{N+1},\vY^c,D_1,D_2)\markov(\hat{\mathbf{M}}_1^N, \hat{\mathbf{M}}_2^{N}). \label{markov-lanc}
\end{align}

Note that in \cite{isit2011} the Markov relation $(\mathbf X_1^N, \mathbf X_2^{N})$ $\markov$ $(\mathbf Y^{N+1},D_1,D_2)$ $\markov$ $(\vY^{N+1},\vY^c)$ was assumed, which need not hold in general. It seems that Poltyrev \cite{Poltyrev} also made this error.

Continuing the estimations (\ref{2})-(\ref{2b}),
\begin{align}
Nn &(R_1+R_2)\leq \I(\vX_1^{N},\vX_2^{N} \wedge \vY^{N+1},\vX^{sw},D_1,D_2 ) + Nn\eps_n\\
 =& \hH(\vX_1^{N},\vX_2^N)+ Nn\eps_n-\hH(\vX_1^{N},\vX_2^N|\vY^{N+1},\vX^{sw},D_1,D_2)\\
 =& \hH(\vX_1^{N},\vX_2^N|D_1,D_2)+ Nn\eps_n -\hH(\vX_1^{N},\vX_2^N|\vY^{N+1},D_1,D_2)\notag\\&+\hH(\vX_1^{N},\vX_2^N|\vY^{N+1},D_1,D_2)-\hH(\vX_1^{N},\vX_2^N|\vY^{N+1},\vX^{sw},D_1,D_2)\label{2b-uj} \\
 =& \I(\vX_1^{N},\vX_2^{N} \wedge \vY^{N+1}|D_1,D_2)+ Nn\eps_n +\I(\vX^{sw} \wedge \vX_1^{N},\vX_2^N|\vY^{N+1},D_1,D_2) \\
 \leq& \hH(\vY^{N+1} |D_1,D_2) - \hH(\vY^{N+1}|\vX_1^{N},\vX_2^{N},D_1,D_2)+4n\log{|\iX|} + Nn\eps_n\\
 =& \hH(\vY^{N+1}|D_1,D_2)+4n\log{|\iX|}+Nn\eps_n\notag\\&-\sum_{j=0}^{N} \sum_{i=0}^{n-1} \hH(Y_{nj+i}|\vY_0^{nj+i-1} \vX_1^N, \vX_2^N,D_1,D_2)\label{2c}\\
 \leq& \sum_{j=0}^{(N+1)n-1}\hH(Y_j|D_1,D_2)+4n\log{|\iX|}+Nn\eps_n \notag\\
 &- \sum_{j=1}^{N-1} \sum_{i=0}^{n-1} \hH(Y_{nj+i}| X_{1,nj+i+D_1}, X_{2,nj+i+D_2},D_1,D_2).\label{2e}
\end{align}

In (\ref{2e}) we dropped some negative terms (notice that $j$ runs from $1$ to $N-1$). Introduce the random variable $\tilde Y_i$ linked to the random variables $X_{1,i\oplus D_1}$,$X_{2,i\oplus D_2}$ by the channel $W$ for all $i\in \{0,1,\dots,n-1\}$, where $\oplus$ denotes the addition modulo $n$. For all $j$ the joint distribution of $(D_1,D_2,X_{1,nj+i+D_1},X_{2,nj+i+D_2},Y_{nj+i})$ is the same as the joint distribution of $(D_1,D_2,X_{1,i\oplus D_1},X_{2,i\oplus D_2},\tilde Y_{i})$. Using this substitution, (\ref{2e}) can be further bounded from above by:
\begin{align}
  \leq& (N-1) \sum_{i=0}^{n-1} \hH(\tilde Y_{i}|D_1,D_2)+2n\log|\irott Y|+4n\log{|\iX|} \notag\\
 &- (N-1)\sum_{i=0}^{n-1} \hH(\tilde Y_i| X_{1,i\oplus D_1} X_{2,i\oplus D_2},D_1,D_2)+Nn\eps_n \label{2d} \\
 =& (N-1)\sum_{i=0}^{n-1} \I(X_{1,i\oplus D_1},X_{2,i\oplus D_2} \wedge \tilde Y_i|D_1,D_2)+Nn\eps_n+4n\log{|\iX|}+2n\log|\irott Y|.
\end{align}

Dividing by $nN$ and introducing the random variable $Q$ uniformly distributed on $\{0,1,\dots,n-1 \}$ and independent of the others we get:

\begin{align}
 R_1&+R_2 \leq\notag\\
\leq& \frac{N-1}{Nn}  \sum_{i=1}^n \I(X_{1,i\oplus D_1},X_{2,i\oplus D_2} \wedge \tilde Y_i|D_1,D_2)+\eps_n+\frac{2\log|\irott Y|}{N}+\frac{4\log|\irott X|}{N}\\
\leq& \frac{N-1}{N} \I(X_{1,Q\oplus D_1},X_{2,Q\oplus D_2} \wedge \tilde Y_Q|Q,D_1,D_2)+\eps_n+\frac{2\log|\irott Y|}{N}+\frac{4\log|\irott X|}{N}
\end{align}
If $N \rightarrow \infty$ then
\begin{align}
 R_1+R_2 &\leq  \I(X_{1,Q\oplus D_1},X_{2,Q\oplus D_2} \wedge \tilde Y_Q|Q,D_1,D_2)+\eps_n.
\end{align}
$\blacksquare$
\begin{Cor} \label{relativedelay}
Under the assumptions of Theorem $\ref{main-converse-prop-alt}$ the following bounds hold in the 2-senders case:
\begin{align}
R_1&\leq \I(X_{1,Q} \wedge \hat{Y}_{Q}|X_{2,Q\ominus D},Q,D)+\eps_n\\
R_2&\leq \I(X_{2,Q\ominus D} \wedge \hat{Y}_{Q}|X_{1,Q},Q,D)+\eps_n\\
R_1+R_2 &\leq \I(X_{1,Q},X_{2,Q\ominus D}\wedge \hat{Y}_{Q}|Q,D)+\eps_n.
\end{align}
Here $Q$ is uniformly distributed on $\{0,1,\dots,n-1\}$ and independent of $D_1$, $D_2$ and the message flows of the senders, $\ominus$ denotes the subtraction modulo $n$, $D=D_1 \ominus D_2$ is the relative delay between the two senders and $\hat{Y}_{Q}$ is linked to $X_{1,Q},X_{2,Q \ominus D}$ through the channel $W$.
\end{Cor}
\emph{Proof:}
Expand the right sides of the equations (\ref{felsobecsles1}),(\ref{felsobecsles2}),(\ref{felsobecsles3}) as sums for the possible values of $Q,D_1,D_2$, e.g.
\begin{align}
& \I(X_{1,Q\oplus D_1},X_{2,Q\oplus D_2}\wedge \tilde Y_{Q}|Q,D_1,D_2)= \notag\\
&= \sum_{q}\sum_{d_1}\sum_{d_2} \frac{1}{n} \cdot \Prob (D_1=d_1) \Prob (D_2=d_2) \I(X_{1,q\oplus d_1},X_{2,q\oplus d_2}\wedge \tilde Y_{q,d_1,d_2}).
\end{align}
Substituting ${q'}=q\oplus d_1$ and $d=d_1 \ominus d_2$, and renaming $q'$ to $q$, the Corollary is proved. $\blacksquare$

\section{Known capacity regions with a new insight}

\subsection{The asynchronous one-sender model} \label{egyfelhasznalo}
In this section the asynchronous model from section 2 is analyzed where there is just one sender ($K=1$). This will provide the basics for the decoding method of the K-AMAC in general.

In case of $K=1$, $W: \mathcal{X} \rightarrow \mathcal{Y}$ denotes a classical DMC. For the sake of clarity, we omit from the notations of Section $2$ the index corresponding to the unique sender. Let $\{ \mathbf{x}(1),\mathbf{x}(2), \dots,  \mathbf{x}(\mathcal{M}) \}$ denote the codewords of the codebook of the sender, where $\mathcal{M}=2^{nR}$ is the number of codewords in the codebook of the sender. The coordinates of $\mathbf{x}(i)$ are denoted by $(x_{0}(i),x_{1}(i), \dots , x_{n-1} (i))$.

The difference between this model and the classical one is that the task of the receiver is not just decoding the codewords but also to find the beginning of the codewords. Note that related problem have been considered in the literature, for example in \cite{erosaszinkron,polyanski}. The known results, however, do not directly apply for our purposes.

\begin{Thm} \label{egy}
For each version of the model the capacity region of the one sender asynchronous model is the same as that of the classical model: $[0, \max_{p} (I(p,W)) ]$, in case of arbitrary delay system.
\end{Thm}
\begin{Rem} \label{delaydet}
It has crucial importance in the proof of Theorems \ref{teljesaszinkron} and \ref{ketto} that in the achievability proof below, beyond decoding the codewords, the receiver also finds out the delay of the sender.
\end{Rem}
\emph{Proof of Theorem \ref{egy}:} The converse part follows from Theorem \ref{main-converse-prop-alt}.

In order to prove the direct part it is enough to restrict attention to uninformed $L$-block decoder and to maximal error; actually $L=1$ suffices. Classical random argument is used. Let $p$ be an arbitrary distribution over the input alphabet $\mathcal{X}$. Chose the symbols of codewords in the codebook of rate $0< R <I(p,W)$ independently according to $p$. Let $P^n(\mathbf{x}^n,\mathbf{y}^n)$ be the joint distribution on $\mathcal{X}^n \times \mathcal{Y}^n$ induced by the n-th power of $p$ and by the memoryless channel $W$. Let $q^n$ be the marginal of $P^n$ on $\mathcal{Y}^n$.
We define the decoder as follows. In order to estimate the 0-th sent message $M_{1,0}$, the receiver first examines the $n$-tuple of outputs $(Y_{-n+1},Y_{-n+2}, \dots, Y_{0})$, then it examines the next $n$-tuple $(Y_{-n+2},Y_{-n+3}, \dots, Y_{1})$, etc. until the $n$-tuple $(Y_{0},Y_{1}, \dots Y_{n-1})$. The estimate will be $\hat{M}_{1,0} = s$ if among the examined $n$-tuples there is a unique one denoted by $Y^n$, for which $((X_0 (s),X_1 (s), \dots, X_{n-1} (s)),Y^n)$ belongs to the typical set
\begin{equation}
S_{n}^{\delta}:=\left\{  (\mathbf{x}^n,\mathbf{y}^n): \left| \frac{1}{n} log \frac{P^n(\mathbf{x}^n,\mathbf{y}^n)}{p^n(\mathbf{x}^n) q^n(\mathbf{y}^n)}  - I(p,W)  \right| \le \delta   \right\},
\end{equation}
and also this $s$ is unique.

It can be assumed that $M_{1,0}$ is fixed, say $M_{1,0}=1$. It is clear from the classical channel coding theorem that if the decoder examines an $n$-tuple $Y^n$ which is the output of the whole codeword $\mathbf{X}(1)$, then the decoder will find $\mathbf{X}(1)$ but no other codewords jointly typical with $Y^n$, with probability exponentially close to $1$. Hence we only have to discuss the cases when the decoder examines output symbols in a window of length $n$, in which one part of the channel input symbols are coming from the $1$st codeword and the other part from another codeword $r$. The probability that $M_{1,-1}$ or $M_{1,1}$ is equal to $1$ is exponentially small. Hence it can be assumed that $r \ne 1$. Furthermore, it can be assumed that the input window starts with the $r'th$ codeword as the opposite case is similar. Hence, the channel input symbols in the examined output window can be written as $(X_{n-l}(r), \dots,X_{n-1}(r),X_{0}(1), \dots,X_{n-l-1}(1))$ for some $r \in  \{ 1, \dots, \mathcal{M} \}, n > l > 0,r \ne 1$.  We will show that the probability of incorrectly recognizing typicality in this window is small. The probability, conditioned on the previously presented structure of the examined window, that the $s$'th codeword will be typical with this examined output $n$ tuple can be written as:
\begin{align}
& \Prob_{ cond} \left\{
\left(   X_1 (s), \dots, X_n (s) ,  Y_{1},\dots, \dots, Y_{n} \right) \in S_{n}^{\delta} \right\}   \notag\\
&=  \sum_{(\mathbf{x}^n (s), \mathbf{y}^n) \in S_{n}^{\delta}} p^{n}(\mathbf{x}^n (s)) \cdot \Prob_{cond} \left\{ (Y_{1}, \dots, Y_{n})=\mathbf{y}^n |(X_1 (s), \dots, X_n (s))=\mathbf{x}^n (s) \right\} \\
&=  \sum_{(\mathbf{x}^n (s), \mathbf{y}^n) \in S_{n}^{\delta}} p^{n}(\mathbf{x}^n (s)) \frac{q^{n}(\mathbf{y}^n)}{q^{n}(\mathbf{y}^n)}  \Prob_{ cond} \left\{ (Y_{1}, \dots, Y_{n})=\mathbf{y}^n |(X_1 (s), \dots, X_n (s))=\mathbf{x}^n (s) \right\}  \\
&\le \sum_{(\mathbf{x}^n (s), \mathbf{y}^n) \in S_{n}^{\delta}} \frac{2^{-n(I(p,W)-\delta)}P^{n}(\mathbf{x}^n (s),\mathbf{y}^n)}{q^{n}(\mathbf{y}^n)} \Prob_{ cond} \left\{ (Y_{1}, \dots, Y_{n})=\mathbf{y}^n |(X_1 (s), \dots, X_n (s))=\mathbf{x}^n (s) \right\}  \\
&=  2^{-n(I(p,W)-\delta)} \sum_{(\mathbf{x}^n (s), \mathbf{y}^n) \in S_{n}^{\delta}} P^{n}(\mathbf{x}^n (s)|\mathbf{y}^n) \Prob_{ cond} \left\{ (Y_{1}, \dots, Y_{n})=\mathbf{y}^n |(X_1 (s), \dots, X_n (s))=\mathbf{x}^n (s) \right\}. \label{utolso}
\end{align}

In the above derivation the definition of the set $S_{n}^{\delta}$ is used, and in the last equation $P^n(\mathbf{x}^n (s)|\mathbf{y}^n)$ denotes the conditional probability induced by the joint distribution $P^n$. At this point we have to use the structure of the examined window. It is known that in this window the second part of the $r$-th codeword and the first part of the $1$st codeword were sent. We should distinguish three cases: $\{ s \ne r, s \ne 1 \}$, $\{ s \ne r, s=1 \}$, $\{ s=r, s \ne 1 \}$. The first case follows from the fact that $\Prob_{ cond} \left\{ (Y_{1}, \dots, Y_{n})=\mathbf{y}^n |(X_1 (s), \dots, X_n (s))=\mathbf{x}^n (s) \right\}$ is equal to $ \Prob_{ cond} \left\{ (Y_{1}, \dots, Y_{n})=\mathbf{y}^n \right\}$. The remaining two cases can be treated very similarly. For the sake of brevity we will demonstrate the case $\{ s \ne r, s=1 \}$ when the expression (\ref{utolso}) is bounded from above by
\begin{align} \label{szumma}
& \left( 2^{-n(I(p,W)-\delta)} \right) \cdot \sum_{\mathbf{x}^{n}(r) \in \mathcal{X}^n} p^n(\mathbf{x}^{n}(r)) \sum_{(\mathbf{x}^n (1), \mathbf{y}^n) \in \mathcal{X}^n \times \mathcal{Y}^n}  \left[ \prod_{h=0}^{n-1} P(x_{h} (1)|y_{h}) \right] \cdot \notag\\
& \cdot \left[ \prod_{h=0}^{l-1} W(y_h | x_{n-l+h}(r))   \right]  \left[ \prod_{h=l}^{n-1} W(y_h | x_{h-l} (1) )  \right].
\end{align}
Sum in the following order: $x_{n-1} (1)$, $y_{n-1}$, $x_{n-2} (1)$, $y_{n-2}$, $\dots$, $x_{l} (1)$, $y_{l}$, we get the following expression:
\begin{align}
& \left( 2^{-n(I(p,W)-\delta)} \right) \cdot \sum_{\mathbf{x}^{n}(r) \in \mathcal{X}^n} p^n(\mathbf{x}^{n}(r)) \sum_{(\mathbf{x}^l (1), y^l) \in \mathcal{X}^l \times \mathcal{Y}^l} \left[ \prod_{h=0}^{l-1} P(x_{h} (1)|y_{h}) \right] \cdot \notag\\
&\cdot \left[ \prod_{h=0}^{l-1} W(y_h | x_{n-l+r}(r))   \right] = 2^{-n(I(P,W)-\delta)}.
\end{align}
Indeed, the inner sum is equal to $1$, because its terms may be regarded as a joint distribution of a memoryless channel with inputs $y_i$ and outputs $x_i (1)$.

In the above derivation we demonstrated that the probability that the receiver finds the s-th codeword typical with an output $n$ tuple whose input symbols consist of two different codewords, can be bounded from above by $2^{-n(I(p,W)-\delta)}$. Using the union bound over all codewords and over all the $n$ tuples examined by the decoder, gives that the probability of recognizing one of the codewords in a window where the inputs are from two different codewords is less then $n \mathcal{M}  2^{-n(I(p,W)-\delta)}$.

The above argument shows that the maximal error over delays of the random code is exponentially small in $n$.  Hence there exists a sequence of deterministic coding-decoding systems with exponentially small maximal error over delays. Optimizing the distribution $P$, we can see that $\max_{p} I(p,W)$ is achievable. $\blacksquare$

\begin{Rem}
Cases $\{ s \ne r, s=1 \}$, $\{ s=r, s \ne 1 \}$ are the main difficulties in this proof. The tricky summation in equation (\ref{szumma}) which solves these difficulties is adopted from Gray \cite{Gray}.
\end{Rem}
\begin{Rem}
Note that using \cite{Gray} a stronger result can be proved: any sequence of deterministic codes  which work well for the classical channel coding model, can be modified to work well for the asynchronous model. Namely, if the same random sync sequence of length $k\approx \log^{2}(n)$ is appended to each of the original codewords, then with probability tending to $1$ it is possible to detect the sync sequence and decode the original codewords. This is how Theorem \ref{egy} was proved in the first version of the paper. The authors are indebted to anonymous reviewers for pointing out that the proof becomes somewhat simpler if no sync sequence is used.
\end{Rem}

\subsection{The totally asynchronous case}\label{totally asynchronous case}
From this point on, the paper strongly relies on the results of \cite{Urbanke} and \cite{rimoldi}. Though the reader is assumed familiar with the concepts of successive decoding and rate splitting, the basics will be summarized below.

Let $W$ be a channel with $K$ senders.
\begin{Def} \label{politop}
The convex polytope $ \mathcal{R} \left[ W; p(x_1,x_2, \dots, x_K)  \right]$ is the set of rate tuples $\mathbf{R} \in (\mathbb{R}^{+})^{K}$ such that
\begin{equation} \label{egyenlotlenseg}
R(S) \le I(\mathbf{X}_{S} \wedge Y|\mathbf{X}_{S^{c}}) \text{     }, S \subseteq \left[ K \right],
\end{equation}
where the joint distribution of $X_1 , X_2 \dots, X_K$ is $p(x_1, x_2 \dots, x_K)$ and $Y$ is connected to  $X_1 ,X_2, \dots, X_K$ by the channel $W$.
\end{Def}
\begin{Def} \label{politop2}
Let $C$ denote the following set:
\begin{equation}\label{C}
C:= \bigcup_{p_{X_1} \times p_{X_2} \times \dots \times p_{X_K} } \mathcal{R} \left[W; p_{X_1} \times p_{X_2} \times \dots \times p_{X_K}  \right]
\end{equation}
where the union is over all product distributions.
\end{Def}
\begin{Def} \label{dominant} The set of rate tuples $(R_1,R_2,\dots, R_K)$ from $ \mathcal{R} \left[W; p_{X_1} \times p_{X_2} \times \dots \times p_{X_K}  \right]$ for which $R(\left[ K \right])=\I(\mathbf{X}_{\left[ K \right]} \wedge Y)$ is the dominant face of $\mathcal{R} \left[W; p_{X_1} \times p_{X_2} \times \dots \times p_{X_K}  \right]$. It is denoted by $D( \mathcal{R} \left[W; p_{X_1} \times p_{X_2} \times \dots \times p_{X_K}  \right])$.
\end{Def}

\begin{Def}
We say that $(R_1,R_2,\dots, R_K)$ is dominated by $(\tilde R_1, \tilde R_2,\dots, \tilde
R_K)$ if $R_1 \leq \tilde R_1$,  $R_2 \leq \tilde R_2$,$\dots$, $R_K \leq \tilde R_K$.
\end{Def}

It can be seen\footnote{\cite{Urbanke} states it as a consequence of the fact that $ \mathcal{R} \left[W; p_{X_1} \times p_{X_2} \times \dots \times p_{X_K}  \right]$ is a polymatroid, which was observed in \cite{matroid1}, \cite{matroid2}.} that the points of $D( \mathcal{R} \left[W; p_{X_1} \times p_{X_2} \times \dots \times p_{X_K}  \right])$ cannot be dominated by other points of $ \mathcal{R} \left[W; p_{X_1} \times p_{X_2} \times \dots \times p_{X_K}  \right]$, but any point from
$ \mathcal{R} \left[W; p_{X_1} \times p_{X_2} \times \dots \times p_{X_K}  \right]$ can be dominated by a point from the dominant face.

\begin{Rem}
According to Definition \ref{kapacitas}, if $(R_1,R_2,\dots,R_K)$ is dominated by an achievable rate vector then the rate vector $(R_1,R_2,\dots,R_K)$ is also achievable.
\end{Rem}

Recall from \cite{Urbanke} the description of the vertices of $D(\mathcal{R} \left[W; p_{X_1} \times p_{X_2} \times \dots \times p_{X_K}  \right])$. Let $\pi=(\pi_{1},\pi_{2}, \dots, \pi_{K})$ be an ordering of $\left[ K \right]$. For all $i \in \left[ K \right]$ let $R^{\pi}_{\pi_{i}}$ be equal to $I(X_{\pi_{i}} \wedge Y|\mathbf{X}_{ \{\pi_{1},\dots,\pi_{i-1}\}})$. For example if $K=3$, and $\pi=(2,3,1)$, then $R_2^{\pi}=I(X_2 \wedge Y)$,  $R_3^{\pi}=I(X_3 \wedge Y| X_2)$, $R_1^{\pi}=I(X_1 \wedge Y| X_2,X_3)$. Then the rate vector $\mathbf{R}^{\pi}=(R^{\pi}_{1},R^{\pi}_{2}, \dots, R^{\pi}_{K})$ is a vertex, and all vertices of $D(\mathcal{R} \left[W; p_{X_1} \times p_{X_2} \times \dots \times p_{X_K}  \right])$ can be written in this way with appropriate $\pi$. Note that the vertices $\mathbf{R}^{\pi}$ need not be all distinct.

In the Appendix of \cite{Urbanke} it is proved for informed $L=K$ block decoder that in the totally asynchronous case $\mathbf{R}^{\pi} \in$ $D(\mathcal{R} \left[W; p_{X_1} \times p_{X_2} \times \dots \times p_{X_K}  \right])$ can be achieved by successive decoding with ordering $\pi$. We summarize the proof for $\mathbf{R}^{\{1,2,\dots, K \}}$. The coding/decoding system is randomly constructed the following way. The symbols of the codebooks of the senders are chosen independently according to the appropriate input distributions. The receiver first decodes by joint typicality the codewords of the first sender, considering the random codewords of the other senders as noise. This means that the receiver behaves as if there were only one sender and the channel was the following:
\begin{equation}
W^{1}(y|x_1)=\sum_{x_2 \in \mathcal{X}_2} \sum_{x_3 \in \mathcal{X}_3} \cdots \sum_{x_k \in \mathcal{X}_K} p_{X_2}(x_2)p_{X_3}(x_3)\cdots p_{X_K}(x_K) W(y|x_1,x_2,\dots, x_K).
\end{equation}
Next the receiver decodes the codewords of the second sender by joint typicality using the already decoded codewords of the first sender, considering the other senders as noise. This means that the receiver behaves as it would in a one sender model when the channel was the following:
\begin{equation}
W^{2}(y,x_1|x_2)=  \sum_{x_3 \in \mathcal{X}_3} \sum_{x_4 \in \mathcal{X}_3} \cdots \sum_{x_k \in \mathcal{X}_K} p_{X_1}(x_1)p_{X_3}(x_3)\cdots p_{X_K}(x_K) W(y|x_1,x_2,\dots, x_K).
\end{equation}
The codewords of the other senders are decoded similarly. In the final decoding step the receiver decodes the codewords of the $K$'th sender by joint typicality using the already decoded codewords of all the other senders. This means that the receiver behaves as it would in a one sender model when the channel was the following:
\begin{equation}
W^{K}(y,x_1,x_2, \dots, x_{K-1}|x_K)= p_{X_1}(x_1)p_{X_2}(x_2)\cdots p_{X_{K-1}}(x_{K-1}) W(y|x_1,x_2,\dots, x_K).
\end{equation}
More detail can be found in the Appendix of \cite{Urbanke}.

Now recall the notion of individual split from \cite{Urbanke} with splitting function $f(x_a,x_b)=max(x_a,x_b)$. A split of sender $i$ with input distribution $p_{X_i}$ on $\mathcal{X}_i = \{0,1,\dots, |\mathcal{X}_{i}| -1 \}$ results in two virtual senders $ia$, $ib$ with distributions $p_{X_{ia}}$ and $p_{X_{ib}}$, also on $\mathcal{X}_i$, explicitly determined by $p_{X_i}$ and a splitting parameter, such that the splitting function $f(x_a,x_b)=max(x_a,x_b)$ maps $p_{X_{ia}} \times p_{X_{ib}}$ into $p_{X_i}$.

Section 2 of \cite{Urbanke} shows in the totally asynchronous case that each  $\mathbf{R} \in D(\mathcal{R} \left[W; p_{X_1} \times p_{X_2} \times \dots \times p_{X_K}  \right])$ can be achieved with Rate Splitting via at most $K-1$ splits\footnote{The stronger result of Section 3 of \cite{Urbanke} is not necessary in this paper.}. This means that a good code for $W$ with rate vector $\mathbf{R}$ can be obtained from a code with successive decoding for an auxiliary channel $W'_{\mathbf{R}}$ with $2K-1$ virtual senders constructed by splitting the original senders, perhaps some of them split repeatedly and others not at all; the rate vector of this code equals the vertex $\mathbf{R'}^{\pi}$ of the dominant face of $\mathcal{R} \left[  W'_{\mathbf{R}}; p_{\tilde{X}_1} \times p_{\tilde{X}_2} \times \cdots \times p_{\tilde{X}_{2K-1}} \right]$ for some ordering $\pi$ and distributions $p_{\tilde{X}_1} \times p_{\tilde{X}_2} \times \cdots \times p_{\tilde{X}_{2K-1}}$. In particular, the $i$'th coordinate of $\mathbf{R}$ is the sum of those coordinates of $\mathbf{R'}^{\pi}$ that correspond to the virtual senders into which the $i$'th sender has been split, $i=1,2,\dots,K$.

\begin{Thm} \label{teljesaszinkron}
In the totally asynchronous case (Example \ref{pelda1}), for each model version the capacity region is $C$.
\end{Thm}
\emph{Proof:} In the converse part it is enough to treat the case of an informed infinite decoder and average error. The right side of eq. (\ref{upperbound}) can be bounded from above as follows.
\begin{align}
&\I(\mathbf X_{S,Q\oplus D_S} \wedge \tilde Y_Q|\mathbf X_{S^c,Q \oplus D_{S^c}},Q,\VD)=\\
&=\hH(\tilde Y_Q | \mathbf X_{S^c,Q \oplus D_{S^c}},Q,\VD) - \hH(\tilde Y_Q | \mathbf X_{\left[ K \right],Q \oplus D_{\left[ K \right]}}, Q,\VD) \\
&=\hH(\tilde Y_Q | \mathbf X_{S^c,Q \oplus D_{S^c}},Q,\VD) - \hH(\tilde Y_Q | \mathbf X_{\left[ K \right],Q \oplus D_{\left[ K \right]}})\label{3a}\\
&\leq \hH(\tilde Y_Q | \mathbf X_{S^c,Q \oplus D_{S^c}}) - \hH(\tilde Y_Q | \mathbf X_{\left[ K \right],Q \oplus D_{\left[ K \right]}})\\
&=\I(\mathbf X_{S,Q\oplus D_S} \wedge \tilde Y_Q|\mathbf X_{S^c,Q \oplus D_{S^c}}) \label{3b}
\end{align}
where (\ref{3a}) comes from the fact that the output depends only on the input variables.

From the fact that the delays are independent and uniform it follows that the random variables $\{ Q\oplus D_i, i \in \left[ K \right] \}$ are independent, hence the random variables $\{ X_{i,Q\oplus D_i}, i \in \left[ K \right] \}$ are also independent. On account of this, the converse statement follows from Theorem \ref{main-converse-prop-alt}.

The achievability part needs one modification of the proof in the Appendix of \cite{Urbanke} of the assertion that $C$ is achievable with informed $L=2K-1$-block decoder, considering maximal error. In order to get rid of the assumption that the delays are known to the receiver, it is enough to use the synchronization method from Subsection \ref{egyfelhasznalo} in the successive steps of achievability of vertices. Note that it is important that in the successive steps the decoder finds the exact delay of the actual sender (see Remark \ref{delaydet}). $\blacksquare$

\begin{Rem} \label{majdnemaszinkron}
In case of two senders Corollary  \ref{relativedelay} leads to a stronger result. If the relative delay $D=D_1 \ominus D_2$ is uniformly distributed on the set $\{0,1,\dots,n-1 \}$ then, for each model version, the capacity region is $C$.
\end{Rem}

\section{Even delays} \label{uniformonevennumbercase}
\begin{figure}[Htb]
 \begin{center}
\includegraphics[width=5.6cm]{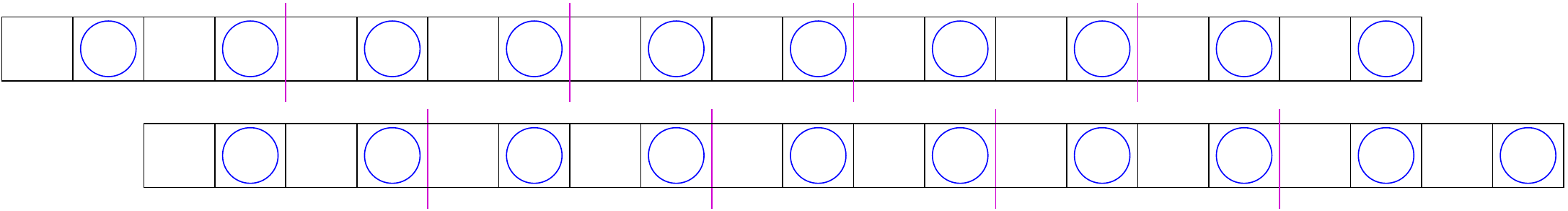}
  \caption{Time sharing when the relative delay is uniform on even numbers}\label{time-share-even}
 \end{center}
\end{figure}

 In \cite{elgamal} an artificial but interesting (from theoretical point of view) model is mentioned as open problem: the possible delays are in the set $\{ 0,1,\dots, \alpha n \}$ for some $\alpha \in (0,1)$. In this section, though this problem is not solved, a similar artificial model is analyzed which also has theoretical interest.
\begin{Thm} \label{parosconverse}
In the even delays case (Example \ref{pelda2}), for each version of the model the capacity region consists of those rate pairs that either belong to $C$ or are linear combinations with weights $\frac{1}{2},\frac{1}{2}$ of points in $C$. Moreover, using coding/decoding systems of odd length, only $C$ can be achieved.
\end{Thm}
\emph{Proof:}
In order to prove the direct part it is enough to restrict attention to uninformed $L$-block decoder and to maximal error. Let $n$ be even. Then the senders can do time sharing with weights $\frac{1}{2},\frac{1}{2}$ using separately the even and the odd symbols and using the coding/decoding method of Theorem \ref{teljesaszinkron}. Figure \ref{time-share-even} demonstrates this fact. Note that in this case $L$ can be chosen as $3$.

In the converse part it is enough to treat the case of an informed infinite decoder and average error. The proof uses Corollary \ref{relativedelay}.

\begin{figure}[Hbt]
 \begin{center}
 \psfrag{R1}{$R_1$} \psfrag{R2}{$R_2$}
  \subfloat[Capacity region in the totally asynchronous case\label{fig-unidelay}]{\includegraphics[width=5.6cm]{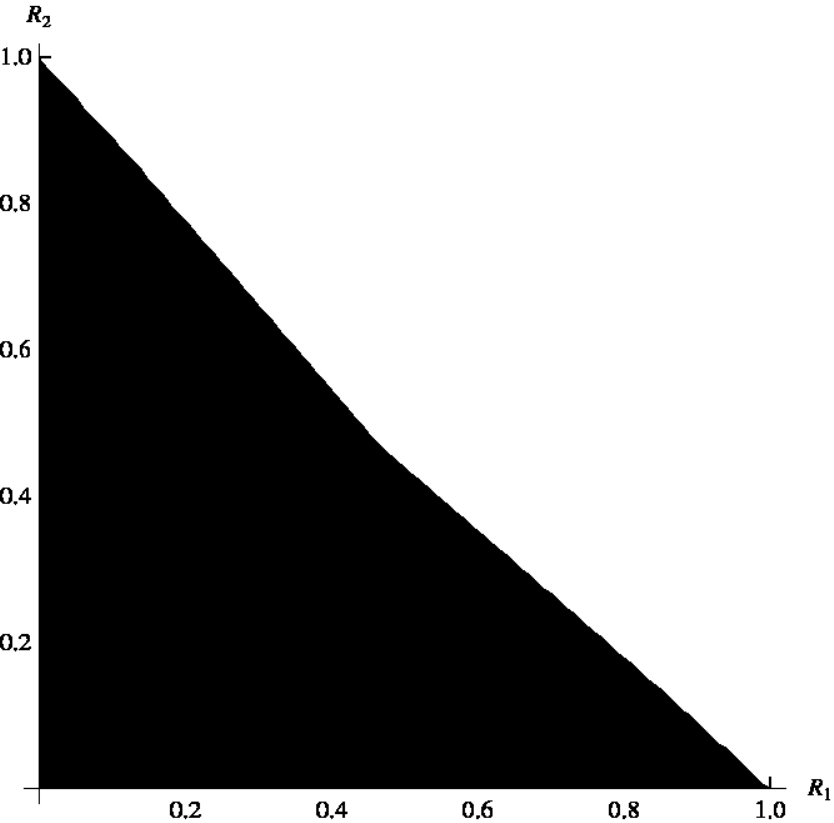}}\hfill
  \subfloat[Capacity region in the even delays case\label{fig-evendelay}]{\includegraphics[width=5.6cm]{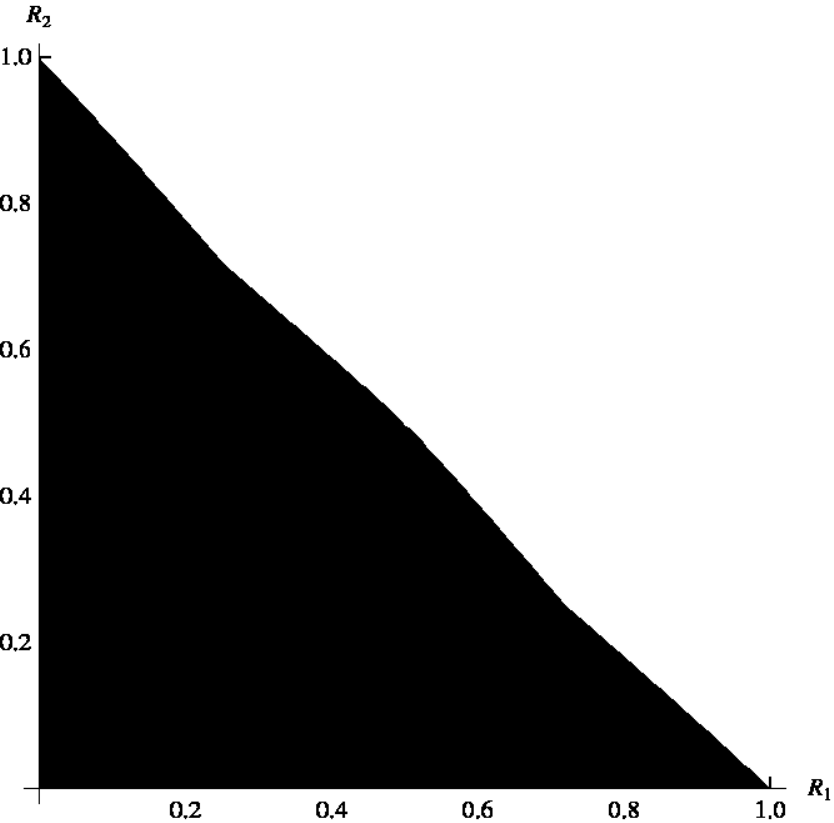}}
  \caption{Change of the capacity region with the change in the distribution of the delays}
 \end{center}
\end{figure}

In case of coding/decoding systems of even length the relative delay is uniformly distributed on the even numbers in $\{ 0, 1, \dots, n-1 \}$. Write the upper bounds in Corollary \ref{relativedelay} as a sum for the possible values of $Q$ and define two random variables $Q_1$, $Q_2$ as uniform on even/odd numbers and independent of each other and everything else. Then the following can be written:
\begin{align}
 R_1\leq& \I(X_{1,Q}\wedge \hat{Y}_Q|X_{2,Q\ominus D},Q,D)+\eps_n\\
 =&\frac{1}{n}\sum_{i=0}^{n-1}\I(X_{1,i}\wedge \hat{Y}_i|X_{2,i\ominus D},D)+\eps_n \\
 \leq&\frac{1}{2}\frac{2}{n}\sum_{i \in \textnormal{odd}}\I(X_{1,i}\wedge \hat{Y}_i|X_{2,i\ominus D})+\frac{1}{2}\frac{2}{n}\sum_{i \in \textnormal{even}} \I(X_{1,i}\wedge \hat{Y}_i|X_{2,i\ominus D})+\eps_n \\
 \leq&\frac{1}{2}\I(X_{1,Q_1}\wedge \hat{Y}_{Q_1}|X_{2,Q_1\ominus D})+\frac{1}{2}\I(X_{1,Q_2}\wedge \hat{Y}_{Q_2}|X_{2,Q_2\ominus D})+\eps_n.
\end{align}
Similarly we get
\begin{align}
 R_2\leq& \frac{1}{2}\I(X_{2,Q_1\ominus D}\wedge \hat{Y}_{Q_1}|X_{1,Q_1})+\frac{1}{2}\I(X_{2,Q_2\ominus D}\wedge \hat{Y}_{Q_2}|X_{1,Q_2})+\eps_n\\
 R_1+R_2\leq& \frac{1}{2}\I(X_{1,Q_1},X_{2,Q_{1}\ominus D}\wedge \hat{Y}_{Q_1})+\frac{1}{2}\I(X_{1,Q_2},X_{2,Q_{2}\ominus D}\wedge \hat{Y}_{Q_2})+\eps_n
\end{align}
where $X_{1,Q_1}$,$X_{2,Q_1\ominus D}$ and $X_{1,Q_2}$,$X_{2,Q_2\ominus D}$ are independent. This proves the converse result for even blocklength (see \cite{Csiszar} Lemma 14.4+, or its generalization, Lemma \ref{imrealt} in Section 6 of this paper).

In the subsequent part of this proof the symbol $n$ denotes odd integer. Now we prove that with coding/decoding systems of odd length, just the union of the pentagons can be achieved. Given such sequence of coding/decoding systems, where $P_e^{n}\rightarrow 0$, let $c_n$ be a sequence with $c_n\to 0$ and $\frac{P_e^{n}}{c_n} \to 0$ such that $c_n n$ is integer.

Recall that the delays $D_1 (n)$ and $D_2(n)$ are independent and uniformly distributed random variables on the set $\{ 0,2,\dots,n-1 \}$. For all $i \in \{ 0,1,\dots,n-1 \}$ let $K(i)$ be the number of those pairs $d_1,d_2 \in \{ 0,2,\dots,n-1 \}$ for which the relative delay $d= d_1 \ominus d_2$ is equal to $i$, then:
\[K(i) =
\left\{  \begin{array}{ll}	\frac{n-i+1}{2}& \textnormal{if\,} i \textnormal{\,is even} \\
				\frac{i+1}{2}	& \textnormal{if\,} i \textnormal{\,is odd.}
         \end{array} \right.
\]

Let $D^{'}_{1} (n)$ and $D'_2(n)$ be two random variables taking values in the set $\{ 0,2,\dots,n-1 \}$ with the following joint distribution. For each $d_1,d_2 \in \{ 0,2,\dots,n-1 \}$, if $d_1 \ominus d_2 \in \{ c_n n, c_n n + 1, \dots, n-1 - c_n n \}$  let $\Prob \left\{ D'_{1} (n)=d_1, D'_{2} (n)=d_2 \right\}$ be equal to $\frac{1}{n(1-2c_n)K(d_1 \ominus d_2)}$, otherwise 0. Then for each $d_1,d_2 \in \{ 0,2,\dots,n-1 \}$ the following bound holds if $n$ is large enough:
\begin{align}
&\frac{4}{(n+1)^2}=\Prob \left\{ D_{1} (n)=d_1, D_{2} (n)=d_2 \right\} \ge c_n \Prob \left\{ D'_{1} (n)=d_1, D'_{2} (n)=d_2 \right\}.
\end{align}
Using the same idea as in the proof of Lemma \ref{alarak} and the fact that $\frac{P_e^{n}}{c_n} \to 0$ we can conclude that the given sequence of coding/decoding systems has average error also tending to $0$ under the delay system $\mathbf{D'}$ described by the random variables $D'_{1} (n)$ and $D'_2(n)$. Hence if we show that under delay system $\mathbf{D'}$ only $C$ can be achieved, the assertion is proved.

Under delay system $\mathbf{D'}$ the relative delay $D' (n) = D'_{1} (n) \ominus D'_2(n)$ is uniformly distributed on the set $\{ c_n n, c_n n+1, \dots, n-1 - c_n n \}$. By Corollary \ref{relativedelay} the following bounds hold for the rates:
\begin{align}
R_1&\leq \I(X_{1,Q} \wedge \hat{Y}_{Q}|X_{2,Q\ominus D'},Q,D')+\eps_n \notag\\
R_2&\leq \I(X_{2,Q\ominus D'} \wedge \hat{Y}_{Q}|X_{1,Q},Q,D')+\eps_n \notag\\
R_1+R_2 &\leq \I(X_{1,Q},X_{2,Q\ominus D'}\wedge \hat{Y}_{Q}|Q,D')+\eps_n.
\end{align}

Let $\bar{D} (n)$ be a random variable uniformly distributed on the set $\{ 0,1,\dots, n-1 \}$. As the variation distance between the product joint distributions of $(Q,D')$ and $(Q,\bar{D})$ tends to $0$, the following differences also tend to $0$ as $n \rightarrow \infty$:
\begin{align}
& \I(X_{1,Q},X_{2,Q\ominus D'}\wedge \hat{Y}_{Q}|Q,D') - \I(X_{1,Q},X_{2,Q\ominus \bar{D}}\wedge \hat{Y}_{Q}|Q,\bar{D}), \notag\\
& \I(X_{1,Q} \wedge \hat{Y}_{Q}|X_{2,Q\ominus D'},Q,D')-\I(X_{1,Q} \wedge \hat{Y}_{Q}|X_{2,Q\ominus \bar{D}},Q,\bar{D}), \notag\\
& \I(X_{2,Q\ominus D'} \wedge \hat{Y}_{Q}|X_{1,Q},Q,D')-\I(X_{2,Q\ominus \bar{D}} \wedge \hat{Y}_{Q}|X_{1,Q},Q,\bar{D}).
\end{align}
Taking into account that $X_{1,Q}$ and $X_{2,Q\ominus \bar{D}}$ are independent, the assertion is proved (See also Remark \ref{majdnemaszinkron}).
$\blacksquare$

\begin{Exa}
There are two well-known examples (\cite{pelda}, \cite{Csiszar}) which show that the convex hull operation can be useful. Here we use \cite{Csiszar}. Let the channel be defined by $\iX_1$ $=\iX_2$ $=\iY$ $=\{0,1\}$, $W(0|0,0)=1$, $W(1|1,0)=W(1|0,1)=1$ and $W(1|1,1)=W(0|1,1)=\frac{1}{2}$. The capacity regions in the totally asynchronous and in the even delays case are shown on Figure \ref{fig-unidelay} and on Figure \ref{fig-evendelay}. In the latter case a hill appears in the middle of the picture.
\end{Exa}

\begin{Rem}
Remarkably, it does not seem possible to extend this result for uniform relative delay distributions on $\{0,1,\dots,n/2\}$, although this distribution has the same entropy as the uniform distribution on even (or odd) numbers. Similar results can be achieved if the distribution is uniform on numbers which are divisible by 3. In this case time sharing with weights $\frac{1}{3},\frac{2}{3}$ becomes possible.
\end{Rem}
\begin{Rem}
This also means that if the senders of the totally asynchronous AMAC want to time share with weights $(\frac{1}{2},\frac{1}{2})$, they can do that if a one-shot 1-bit side-information about the delays is available to the senders.
\end{Rem}

\section{Partly asynchronous three-senders case}

In this section we will prove coding theorem in case of $K=3$, when $D_1 (n)=D_2 (n)$ and $D_3 (n)$ are independent and uniformly distributed on the set $\{ 0, 1, \dots, n-1  \}$.
\begin{Thm} \label{ketto}
In the partly asynchronous three senders case (Example \ref{pelda3}), for each version of the model the capacity region is
\begin{equation}
\bigcup_{p_{X_3}} Conv \left( \bigcup_{p_{X_1} \times p_{X_2}} \mathcal{R} \left[W; p_{X_1} \times p_{X_2} \times p_{X_3}  \right] \right).
\end{equation}
In words, it consists of the convex combination of rate triples from $C$ whose corresponding convex polytopes are defined by the same third distribution. \end{Thm}
\begin{Rem} \label{elegharom}
Using the Carath\'eodory-Frenchel Theorem (e.g. Chapter 15 of \cite{Csiszar}) in Theorem \ref{ketto}, it suffices to take convex combinations involving at most three rate triples.
\end{Rem}
\emph{Proof of the converse part in Theorem \ref{ketto}:}
It is enough to treat the case of an informed infinite decoder and average error. Theorem \ref{main-converse-prop-alt} can be used as follows.

If $S \subset \{ 1,2,3 \}$ then the following bound holds:
 \begin{align}
  R(S)\leq& \sum_{i=1}^n \frac{1}{n} \I(\mathbf{X}_{S,i\oplus D_S} \wedge Y_i|\mathbf{X}_{S^c,i \oplus D_{S^c}},\VD)+\varepsilon_{n}.
 \end{align}
 Summing over the possible values of $D_1 = D_2$ we get the following bounds:
 \begin{align}
  R_1 \leq& \sum_{i=0}^{n-1} \sum_{d=0}^{n-1} \frac{1}{n}  \frac{1}{n} \I(X_{1,i\oplus d} \wedge \tilde{Y}_i | X_{2,i\oplus d},X_{3,i\oplus D_3},D_3) +\varepsilon_{n} \notag\\
  R_2 \leq& \sum_{i=0}^{n-1} \sum_{d=0}^{n-1} \frac{1}{n}  \frac{1}{n}  \I(X_{2,i\oplus d} \wedge \tilde{Y}_i | X_{1,i\oplus d},X_{3,i\oplus D_3},D_3) +\varepsilon_{n} \notag\\
  R_3 \leq& \sum_{i=0}^{n-1} \sum_{d=0}^{n-1} \frac{1}{n}  \frac{1}{n}  \I(X_{3,i\oplus D_3} \wedge \tilde{Y}_i | X_{1,i\oplus d},X_{2,i\oplus d},D_3) + \varepsilon_{n} \notag \\
  R_1+R_2 \leq& \sum_{i=0}^{n-1} \sum_{d=0}^{n-1} \frac{1}{n}  \frac{1}{n}  \I(X_{1,i\oplus d},X_{2,i\oplus d} \wedge \tilde{Y}_i | X_{3,i\oplus D_3},D_3) + \varepsilon_{n} \notag\\
  R_2+R_3 \leq& \sum_{i=0}^{n-1} \sum_{d=0}^{n-1} \frac{1}{n}  \frac{1}{n}  \I(X_{2,i\oplus d},X_{3,i\oplus D_3} \wedge \tilde{Y}_i | X_{1,i\oplus d},D_3) + \varepsilon_{n} \notag\\
  R_1+R_3 \leq& \sum_{i=0}^{n-1} \sum_{d=0}^{n-1} \frac{1}{n}  \frac{1}{n} \I(X_{1,i\oplus d},X_{3,i\oplus D_3} \wedge \tilde{Y}_i | X_{2,i\oplus d},D_3) +\varepsilon_{n} \notag\\
  R_1+R_2+R_3 \leq& \sum_{i=0}^{n-1} \sum_{d=0}^{n-1} \frac{1}{n}  \frac{1}{n}  \I(X_{1,i\oplus d},X_{2,i\oplus d},X_{3,i\oplus D_3} \wedge \tilde{Y}_i| D_3) +\varepsilon_{n}.
 \end{align}

Note that, $X_{3,i\oplus D_3}$ is independent of the other variables, and has the same distribution for all $i$. Note also that the above inequalities can be overestimated by dropping $D_3$ from the condition (same argument as in Theorem \ref{teljesaszinkron}). Hence the converse part follows from Lemma \ref{imrealt} below. $\blacksquare$

The achievability part in Theorem \ref{ketto} is proved later in this section.

\begin{lemma} \label{imrealt}
Given $k$ sets $\mathcal{R} \left[W; p_{X^{i}_1} \times p_{X^{i}_2} \times p_{X^{i}_3}  \right]$, $i \in \left[ k \right]$, a vector $(R_1, R_2, R_3)$ equals a convex combination with weights $\alpha_i$ of $k$ vectors from these sets if and only if they are contained in $\mathcal{R}(\alpha_1,\alpha_2, \dots, \alpha_k)$ which is defined by the following inequalities:
\begin{align}
&0 \le R_1 \le \sum_{i=1}^{k} \alpha_{i} I(X^{i}_{1} \wedge Y^{i} | X^{i}_{2},X^{i}_{3})  \notag\\
&0 \le R_2 \le \sum_{i=1}^{k} \alpha_{i} I(X^{i}_{2} \wedge Y^i | X^{i}_{1},X^{i}_{3})  \notag\\
&0 \le R_3 \le \sum_{i=1}^{k} \alpha_{i} I(X^{i}_{3} \wedge Y^i | X^{i}_{1},X^{i}_{2})  \notag\\
&R_1+R_2 \le \sum_{i=1}^{k} \alpha_{i} I(X^{i}_{1}, X^{i}_{2} \wedge Y^i |X^{i}_{3})  \notag\\
&R_1+R_3 \le \sum_{i=1}^{k} \alpha_{i} I(X^{i}_{1}, X^{i}_{3} \wedge Y^i |X^{i}_{2})  \notag\\
&R_2+R_3 \le \sum_{i=1}^{k} \alpha_{i} I(X^{i}_{2}, X^{i}_{3} \wedge Y^i |X^{i}_{1})  \notag\\
&R_1+R_2+R_3 \le \sum_{i=1}^{k} \alpha_{i} I(X^{i}_{1},X^{i}_{2}, X^{i}_{3} \wedge Y^i). \label{konv}
\end{align}
\end{lemma}
\emph{Proof:} This proof follows the proof of Lemma 14.4+ in \cite{Csiszar}.
The sets $\mathcal{R} \left[W; p_{X^{i}_1} \times p_{X^{i}_2}  \times p_{X^{i}_3}  \right]$, $i \in \left[ k \right]$, and the set $\mathcal{R}(\alpha_1,\alpha_2, \dots, \alpha_k)$ are convex polytopes with 16 vertices. Using the fact that the mutual and the conditional mutual information are always non-negative, it can be easily derived that there are no redundant inequalities between the defining equations of the sets $\mathcal{R} \left[W; p_{X^{i}_1} \times p_{X^{i}_2}  \times p_{X^{i}_3}  \right]$, $i \in \left[ k \right]$, and $\mathcal{R}(\alpha_1,\alpha_2, \dots, \alpha_k)$. This means for example that it is not possible that the sum of the bounds for $R_1+R_2$ and $R_3$ is strictly less then the bound for $R_1+R_2+R_3$. Using this fact the vertices of $\mathcal{R} \left[W; p_{X^{i}_1} \times p_{X^{i}_2}  \times p_{X^{i}_3}  \right]$, $i \in \left[ k \right]$, can be written in the following way. First $\mathbf{v}_{i}^{0}=(0,0,0)$ is a vertex. The remaining $15$ vertices can be divided into three groups of equal size. The first group consists of those vertices $(R_1, R_2, R_3)$ for which $R_1$ is equal to its own bound, i.e, of the vertices:
\begin{align}
\mathbf{v}_{i}^{1}=(I(X^{i}_{1} \wedge Y^{i} | X^{i}_{2},X^{i}_{3}),0,0) \notag\\
\mathbf{v}_{i}^{2}=(I(X^{i}_{1} \wedge Y^{i} | X^{i}_{2},X^{i}_{3}),I(X^{i}_{2} \wedge Y^{i} |X^{i}_{3},0) \notag\\
\mathbf{v}_{i}^{3}=(I(X^{i}_{1} \wedge Y^{i} | X^{i}_{2},X^{i}_{3}),I(X^{i}_{2} \wedge Y^{i} |X^{i}_{3},I(X^{i}_{3} \wedge Y^{i}) \notag\\
\mathbf{v}_{i}^{4}=(I(X^{i}_{1} \wedge Y^{i} | X^{i}_{2},X^{i}_{3}),0,I(X^{i}_{3} \wedge Y^{i} |X^{i}_{2}) \notag\\
\mathbf{v}_{i}^{5}=(I(X^{i}_{1} \wedge Y^{i} | X^{i}_{2},X^{i}_{3}),I(X^{i}_{2} \wedge Y^{i}),I(X^{i}_{3} \wedge Y^{i} |X^{i}_{2})
\end{align}
The two other groups $(\mathbf{v}_{i}^{6}, \mathbf{v}_{i}^{7}, \dots, \mathbf{v}_{i}^{10})$ and $(\mathbf{v}_{i}^{11}, \mathbf{v}_{i}^{12}, \dots, \mathbf{v}_{i}^{15})$  are obtained similarly. 

Note that $\mathcal{R} \left[W; p_{X^{i}_1} \times p_{X^{i}_2}  \times p_{X^{i}_3}  \right]$ can be degenerate in the sense that these sixteen vertices need not be all distinct. The vertices of $\mathcal{R}(\alpha_1,\alpha_2, \dots, \alpha_k)$ are the points $\sum_{i=1}^{k} \alpha_{i} \mathbf{v}_{i}^{j}$, $0 \le j \le 15$. As these vertices are contained in the (convex) set of convex combinations with weights $\alpha_{i}$ of vectors in the sets $\mathcal{R} \left[W; p_{X^{i}_1} \times p_{X^{i}_2} \times p_{X^{i}_3}  \right]$, $i \in \left[ k \right]$, therefore whole $\mathcal{R}(\alpha_1,\alpha_2,\dots,\alpha_k)$ is contained. The reverse inclusion is obvious.
$\blacksquare$

By Definition \ref{dominant}, the points $(R_1,R_2,R_3)$ of $D(\mathcal{R} \left[W; p_{X_1} \times p_{X_2} \times p_{X_3}  \right])$ satisfy the inequalities in (\ref{egyenlotlenseg}), with $R_1 + R_2 + R_3 = I(X_1, X_2, X_3 \wedge Y)$. An edge of this dominant face is characterized by another inequality in (\ref{egyenlotlenseg}) fulfilled with equality. The set $S$ corresponding to that inequality will be called the type of this edge.

The following lemma states that rate triples lying on edges with a fixed type behave similarly in context of rate splitting and successive decoding. It can be considered as a remark to the general theory of \cite{Urbanke}, \cite{rimoldi} in the special case $K=3$.
\begin{lemma} \label{kulcslemma1}
For every fixed nonempty $S \subsetneq \{ 1,2,3 \}$ there exists a 4-senders channel $W'$ derived from $W$ by splitting the first or the second sender, and an ordering $\pi=(\pi_1,\pi_2,\pi_3,\pi_4)$ of the 4 senders with the following property. If $W'$ is derived from $W$ by splitting the first sender, then to any input distributions $p_{X_1}, p_{X_2}, p_{X_3}$ of $W$ and for all $(R_1,R_2,R_3) \in$  $D(\mathcal{R} \left[W; p_{X_1} \times p_{X_2} \times p_{X_3}  \right])$ lying on the edge of type $S$, there exist input distributions $p_{X_{1a}}, p_{X_{1b}}$ and non-negative numbers $R_{1a},R_{1b}$ with $R_{1a}+R_{1b}=R_1$ such that $(R_{1a},R_{1b}, R_2,R_3)$ is the vertex of $D(\mathcal{R} \left[W'; p_{X_{1a}} \times p_{X_{1b}} \times p_{X_2} \times p_{X_3}  \right])$ described by ordering $\pi$. If $W'$ is derived from $W$ by splitting the second sender, then to any input distributions $p_{X_1}, p_{X_2}, p_{X_3}$ of $W$ and for all $(R_1,R_2,R_3) \in$  $D(\mathcal{R} \left[W; p_{X_1} \times p_{X_2} \times p_{X_3}  \right])$ lying on the edge of type $S$, there exist input distributions $p_{X_{2a}}, p_{X_{2b}}$ and non-negative numbers $R_{2a},R_{2b}$ with $R_{2a}+R_{2b}=R_2$ such that $(R_{1},R_{2a}, R_{2b}, R_3)$ is the vertex of $D(\mathcal{R} \left[W'; p_{X_{1}} \times p_{X_{2a}} \times p_{X_{2b}} \times p_{X_3}  \right])$ described by ordering $\pi$.
\end{lemma}
\emph{Proof:}
Assume for example that a rate triple $(R_1,R_2,R_3) \in$
$D(\mathcal{R} \left[W; p_{X_1} \times p_{X_2} \times p_{X_3}  \right])$ lies on the edge of type $S= \{ 1,3 \}$, hence $R_1 + R_2 + R_3 = I(X_1, X_2, X_3 \wedge Y)$, $R_1 + R_3 = I(X_1,X_3 \wedge Y | X_2)$. The other cases are similar. Then $R_2 = I(X_2 \wedge Y)$ and $(R_1, R_3)$ lies on $D(\mathcal{R} \left[\hat{W}; p_{X_1} \times p_{X_3}  \right])$, where $\hat{W}(y,x_2|x_1, x_3)= p_{X_2} (x_2) W(y|x_1, x_2, x_3)$ (see \cite{rimoldi}, beginning of section 3c). Denote by $\hat{W}'$ the three senders channel derived from $\hat{W}$ by splitting the first sender. We could split the third sender instead of the first sender, but we want to leave the third sender unsplit. Using the basic rate splitting result of \cite{Urbanke} for two senders channels, there exist input distributions $p_{X_{1a}},p_{X_{1b}}$ and non-negative numbers $R_{1a},R_{1b}$ with $R_{1a}+R_{1b}=R_1$ such that $(R_{1a},R_{1b},R_3)$ is the vertex of $D(\mathcal{R} \left[\hat{W}'; p_{X_{1a}} \times p_{X_{1b}} \times p_{X_3}  \right])$ described by the ordering $(1a,3,1b)$. Hence, $(R_{1a},R_{1b},R_2, R_3)$ is the vertex of $D(\mathcal{R} \left[W'; p_{X_{1a}} \times p_{X_{1b}} \times p_{X_2} \times p_{X_3}  \right])$ described by the ordering $(2,1a,3,1b)$, where $W'$ is the 4-senders channel derived from $W$ by splitting the first sender. This argument shows that for $S= \{ 1,3 \}$, the channel $W'$ derived from $W$ by splitting the first sender, and the ordering $(2,1a,3,1b)$ on the senders of $W'$ fulfill the requirements of this lemma. $\blacksquare$

The next lemma shows that each $\Vx$ which is not in $C$ but can be written as the convex combination of rate triples from $C$, can be dominated by a convex combination of rate triples from $C$ which lie on edges of same type.
\begin{lemma} \label{kulcslemma2}
Given $k$ sets $\mathcal{R} \left[W; p_{X^{i}_1} \times p_{X^{i}_2} \times p_{X^{i}_3}  \right]$, $i \in \left[ k \right]$, if a vector $\Vx$ is not in $C$, but can be written as $\Vx = \sum_{i=1}^{k} \alpha_i \Vx_i$, where $\Vx_i \in \mathcal{R} \left[W; p_{X^{i}_1} \times p_{X^{i}_2} \times p_{X^{i}_3}  \right]$, $0 \le \alpha_i < 1, i \in \left[ k \right]$, $\sum_{i=1}^{k} \alpha_i=1$, then $\Vx$ can be dominated by an $\Vx'$ which can be written as $\sum_{i=1}^{k} \alpha'_i \Vx'_i$, where $\Vx'_i \in D(\mathcal{R} \left[W; p_{X^{i}_1} \times p_{X^{i}_2} \times p_{X^{i}_3}  \right])$, $0 \le \alpha'_i < 1, i \in i \in \left[ k \right]$, $\sum_{i=1}^{k} \alpha'_i=1$ and the vectors $\Vx'_i, i \in \left[ k \right],$ lie on edges of same type.
\end{lemma}
\emph{Proof:}
It can be assumed that $\alpha_i >0$ for all $i$. If $\Vx_i$ is not on  $D(\mathcal{R} \left[W; p_{X^{i}_1} \times p_{X^{i}_2} \times p_{X^{i}_3}  \right])$ then we can take a dominating $\tilde
\Vx_i$ from the dominant face, for all $i$. Then $\tilde \Vx=\sum_{i=1}^{k} \alpha_i \tilde
\Vx_i$ dominates $\Vx$. So it can be assumed that the rate triple $\Vx_i$ is on $D(\mathcal{R} \left[W; p_{X^{i}_1} \times p_{X^{i}_2} \times p_{X^{i}_3}  \right])$ for all $i$.

\begin{figure}[Htb]
 \begin{center}
\includegraphics[width=5.6cm]{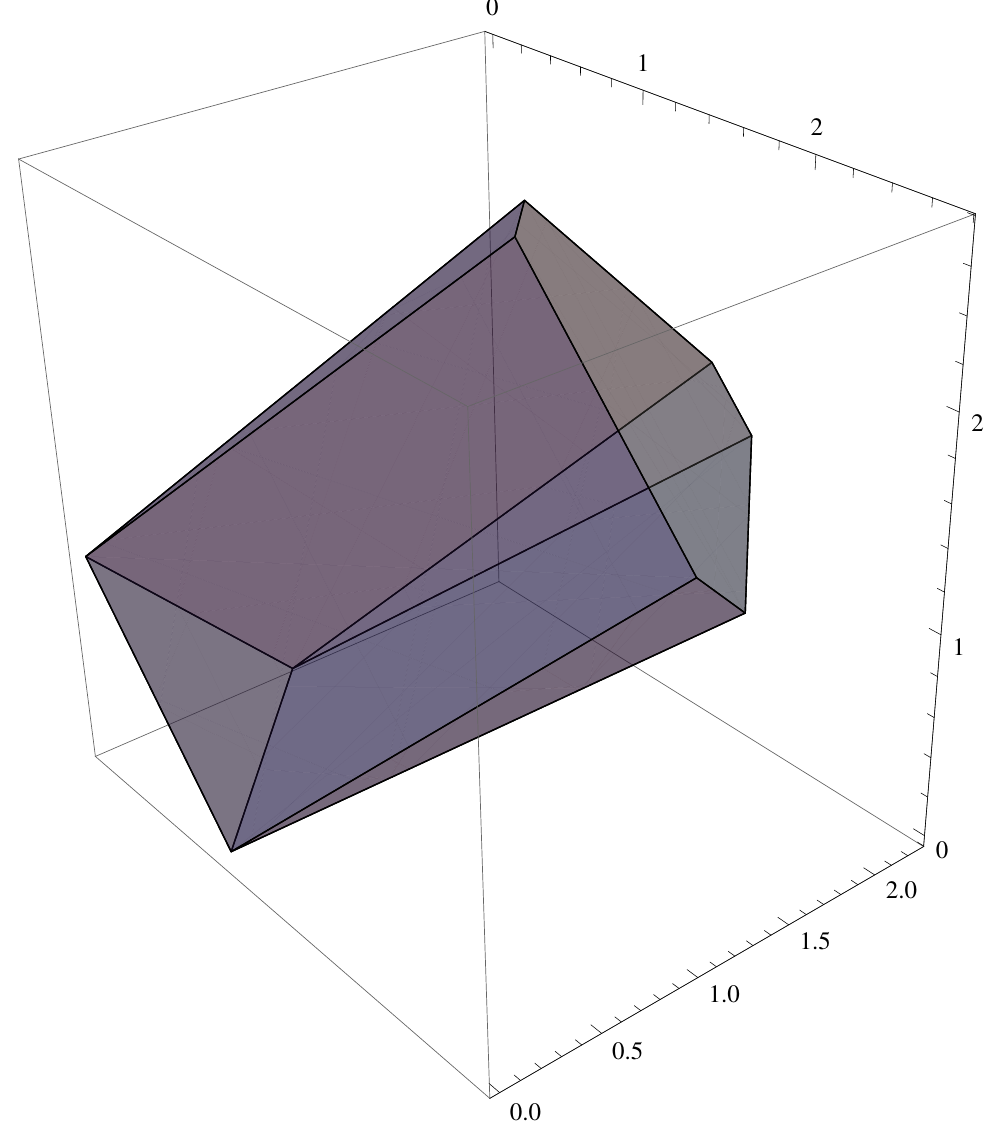}
  \caption{The set of convex combination of two dominant faces. One of them is degenerate (triangle).}\label{polygon}
 \end{center}
\end{figure}

The dominant face of a set $\mathcal{R} \left[W; p_{X^{i}_1} \times p_{X^{i}_2} \times p_{X^{i}_3}  \right]$ is a hexagon\footnote{The hexagon can be degenerated since some vertices can be identical} on a plane with normal vector $(1,1,1)$.
We say that the height of the plane with normal vector $(1,1,1)$ is $a$ if its equation is $x+y+z = a$. The height of a dominant face is the height of its plane. As in Lemma \ref{imrealt} let us consider the set $\mathcal{R}(\alpha_1, \alpha_2, \dots, \alpha_k)$. This is the set of convex combinations with weights $\alpha_i , 1 \le i \le k$ of the sets $\mathcal{R} \left[W; p_{X^{i}_1} \times p_{X^{i}_2} \times p_{X^{i}_3}  \right], 1 \le i \le k$. The dominant face $\mathcal{D}(\alpha_1,\alpha_2, \dots, \alpha_k)$ of $\mathcal{R}(\alpha_1, \alpha_2 ,  \dots, \alpha_k)$ consists of those points $(R_1, R_2, R_3)$ for which $R_1+R_2+R_3 = \sum_{i=1}^{k} \alpha_{i} I(X^{i}_{1},X^{i}_{2}, X^{i}_{3} \wedge Y^i)$. Note that $\Vx \in \mathcal{D}(\alpha_1,\alpha_2 ,  \dots, \alpha_k)$ because the points $\Vx_i$ are on the dominant face of  $\mathcal{R} \left[W; p_{X^{i}_1} \times p_{X^{i}_2} \times p_{X^{i}_3}  \right]$ respectively. Any edge of $\mathcal{D}(\alpha_1,\alpha_2 , \dots, \alpha_k)$ consists of those points for which one of the inequalities (\ref{konv}) is fulfilled with equality. Hence the edges of $\mathcal{D}(\alpha_1, \alpha_2 , \dots, \alpha_k)$ consist of points which are convex combinations with weights $\alpha_1,\alpha_2 ,  \dots, \alpha_k$ of points lying on edges of same type. If $\Vx$ is on an edge of $\mathcal{D}(\alpha_1,\alpha_2, \dots, \alpha_k)$ then we proved the assertion. Hence it can be assumed that $\Vx$ is an inner point of $\mathcal{D}(\alpha_1,\alpha_2 ,  \dots, \alpha_k)$. Suppose first that there exists $m,l$ such that  $I(X_{1}^{m},X_{2}^{m},X_{3}^{m} \wedge Y^{m}) > I(X_{1}^{l},X_{2}^{l},X_{3}^{l} \wedge Y^{l})$. Let us define new weights: If $i \ne m, i \ne l$ then let $\alpha^{'}_{i}=\alpha_{i}$, and let $\alpha^{'}_{m}=\alpha_{m}+ \varepsilon$, $\alpha^{'}_{l}=\alpha_{l} -\varepsilon$. Then the height of $\mathcal{D}(\alpha^{'}_1,\alpha^{'}_2, \dots, \alpha^{'}_k)$ is larger than the height of $\mathcal{D}(\alpha_1,\alpha_2 , \dots, \alpha_k)$. If $\varepsilon$ is small then one of the points of $\mathcal{D}(\alpha^{'}_1,\alpha^{'}_2 , \dots, \alpha^{'}_k)$ will dominate $\Vx$. We increase $\varepsilon$ until this property holds or until $\alpha^{'}_{l}$ becomes $0$. Then, using continuity, an edge point of $\mathcal{D}(\alpha^{'}_1,\alpha^{'}_2 ,  \dots, \alpha^{'}_k)$ will dominate $\Vx$ or $\alpha^{'}_{l}=0$ holds. This argument shows that it is enough to restrict attention to the case when $I(X_{1}^{m},X_{2}^{m},X_{3}^{m} \wedge Y^{m})= I(X_{1}^{l},X_{2}^{l},X_{3}^{l} \wedge Y^{l})$ for all $m,l$. This means that the dominant faces of sets $\mathcal{R} \left[W; p_{X^{i}_1} \times p_{X^{i}_2} \times p_{X^{i}_3}  \right]$ are in the same plane. Using again the continuous change of $\mathcal{D}(\alpha_1,\alpha_2 , \dots, \alpha_k)$: if $\alpha_i \rightarrow 1$, and $\alpha_{j} \rightarrow 0$ if $j \ne i$, then $\mathcal{D}(\alpha_1,\alpha_2 , \dots, \alpha_k)$ tends to $D(\mathcal{R} \left[W; p_{X^{i}_1} \times p_{X^{i}_2} \times p_{X^{i}_3}  \right])$. As $\Vx$ is not in $C$, it is not in $D(\mathcal{R} \left[W; p_{X^{i}_1} \times p_{X^{i}_2} \times p_{X^{i}_3}  \right])$ for either $i \in \left[ k \right]$, hence there are weights $\alpha^{*}_1,\alpha^{*}_2 , \dots, \alpha^{*}_k$ for which $\Vx$ is on an edge of $\mathcal{D}(\alpha^{*}_1,\alpha^{*}_2 , \dots, \alpha^{*}_k)$. So it is a convex combination of points lying on edges of same type.
$\blacksquare$

\emph{The proof of achievability in Theorem \ref{ketto}:}
In order to prove the direct part it is enough to restrict attention to uninformed $L$-block decoder and to maximal error.

Theorem \ref{teljesaszinkron} shows that the rate triples of $C$ can be achieved in the totally asynchronous case considering uninformed $L=5$-block decoder with maximal error. It follows that in the party asynchronous three senders case, $C$ is also achievable considering uninformed $L=5$-block decoder with maximal error with the same coding/decoding method. 

Hence, using also Remark \ref{elegharom}, it is enough to consider points which are not in $C$ but can be written as the convex combination of two or three rate triples from $C$ whose corresponding sets $\mathcal{R} \left[W; p_{X_1} \times p_{X_2} \times p_{X_3}  \right]$ have the same third distribution. Note that the following part of this proof shows that in the latter cases $L=4$-block decoder suffices. For the sake of clarity we deal only with points which can be written as the convex combination of two rate triples whose corresponding sets $\mathcal{R} \left[W; p_{X_1} \times p_{X_2} \times p_{X_3}  \right]$ have the same third distribution. The case of convex combination of three rate triples can be derived similarly.

Let $(R_1, R_2, R_3)$ be in $D(\mathcal{R} \left[W; p_{X_1} \times p_{X_2} \times p_{X_3}  \right])$ and $(\tilde{R}_1, \tilde{R}_2, \tilde{R}_3)$ in  $D(\mathcal{R} \left[W; p_{\tilde{X}_1} \times p_{\tilde{X}_2} \times p_{X_3}  \right])$. Note that the third input distribution is the same in case of both convex polytopes. We want to show that $\alpha (R_1, R_2, R_3)+(1-\alpha)(\tilde{R}_1, \tilde{R}_2, \tilde{R}_3)$  can be achieved in the partially asynchronous three senders case, $\alpha \in (0,1)$. Using Lemma \ref{kulcslemma2} it can be assumed that $(R_1, R_2, R_3)$ and $(\tilde{R}_1, \tilde{R}_2, \tilde{R}_3)$ lie on edges of same type. Without loss of generality it can be assumed that this common type is $S=\{ 1,3 \}$.

Let $W'$ and $\pi$ be the 4-senders channel and the ordering in Lemma \ref{kulcslemma1} for $S=\{1,3\}$. From the proof of Lemma \ref{kulcslemma1} it can be seen that $W'$ is the first sender split version of $W$ and $\pi=(2,1a,3,1b)$. As a consequence of Lemma \ref{kulcslemma1} there exist $R_{1a},R_{2a}$ with $R_{1a}+R_{2a}=R_1$ and $\tilde{R}_{1a},\tilde{R}_{2a}$ with $\tilde{R}_{1a}+\tilde{R}_{2a}=\tilde{R}_1$ and input distributions $p_{X_{1a}}, p_{X_{1b}},p_{\tilde{X}_{1a}}, p_{\tilde{X}_{1b}}$ such that $(R_{1a},R_{1b}, R_2,R_3)$ and $(\tilde{R}_{1a},\tilde{R}_{1b}, \tilde{R}_2,\tilde{R}_3)$  are those vertices of $D(\mathcal{R} \left[W'; p_{X_{1a}} \times p_{X_{1b}} \times p_{X_2} \times p_{X_3}  \right])$ and $D(\mathcal{R} \left[W'; p_{\tilde{X}_{1a}} \times p_{\tilde{X}_{1b}} \times p_{\tilde{X}_2} \times p_{X_3}  \right])$ respectively, which can be described by ordering $\pi$.

If a sender is split, then the delays of the two virtual senders are equal to the delay of the original sender. Hence it is enough to prove that $\alpha(R_{1a},R_{1b}, R_2,R_3)+(1-\alpha)(\tilde{R}_{1a},\tilde{R}_{1b}, \tilde{R}_2,\tilde{R}_3)$ can be achieved for channel $W'$ when the delay system is the following: $D_{1a}(n)=D_{1b}(n)=D_{2}(n)$ and $D_{3} (n)$ are independent and uniformly distributed on the set $\{0,1,\dots, n-1 \}$.

Note that the coordinates of the 4-tuple $(R_{1a},R_{1b}, R_2,R_3)$ can be described as follows: $R_{2}=I(X_{2} \wedge Y)$, $R_{1a}=I(X_{1a} \wedge Y | X_{2})$, $R_{3}=I(X_{3} \wedge Y|X_{1a},X_{2})$ and $R_{1b}=I(X_{1b} \wedge Y|X_{1a},X_{2},X_{3})$, where the joint distribution of $(X_{1a}$, $X_{1b}$, $X_2$, $X_3$, $Y$) is determined by the product input distribution $p_{X_{1a}} \times p_{X_{1b}} \times p_{X_2} \times p_{X_3}$ and the channel transition $W'$. Similarly the coordinates of the 4-tuple $(\tilde{R}_{1a},\tilde{R}_{1b}, \tilde{R}_2,\tilde{R}_3)$ can be described by the equations: $\tilde{R}_{2}=I(\tilde{X}_{2} \wedge \tilde{Y})$, $\tilde{R}_{1a}=I(\tilde{X}_{1a} \wedge \tilde{Y} | \tilde{X}_{2})$, $\tilde{R}_{3}=I(X_{3} \wedge \tilde{Y}|\tilde{X}_{1a},\tilde{X}_{2})$ and $\tilde{R}_{1b}=I(\tilde{X}_{1b} \wedge \tilde{Y}|\tilde{X}_{1a},\tilde{X}_{2},X_{3})$, where the joint distribution of $(\tilde{X}_{1a} , \tilde{X}_{1b} , \tilde{X}_2 , X_3, \tilde{Y})$ is determined by the product input distribution $p_{\tilde{X}_{1a}} \times p_{\tilde{X}_{1b}} \times p_{\tilde{X}_2} \times p_{X_3}$ and the channel transition $W'$.

Random coding argument is used, assuming without any loss of generality that $\alpha n$ and $(1-\alpha)n$ are integers. The symbols of the random codebooks are independent but not identically distributed random variables. The codewords of the virtual senders $1a$, $1b$, and sender $2$ consist of two parts. The first $\alpha n$ symbols have distributions $p_{X_{1a}}$, $p_{X_{1b}}$ and $p_{X_2}$ respectively, while the last $(1-\alpha)n$ symbols have distributions $p_{\tilde{X}_{1a}}$, $p_{\tilde{X}_{1b}}$ and $p_{\tilde{X}_2}$ respectively. The symbols of codewords of sender $3$ are identically distributed according to the distribution $p_{X_3}$. We show that with this codebook structure it is possible to achieve the rate tuple $\alpha(R_{1a},R_{1b}, R_2,R_3)+(1-\alpha)(\tilde{R}_{1a},\tilde{R}_{1b}, \tilde{R}_2,\tilde{R}_3)$, by successive decoding with ordering $(2,1a,3,1b)$ for channel $W'$ if senders $1a$, $1b$, $2$ are synchronized but sender $3$ is not synchronized with them.

Note that we do not assume that the receiver knows the delays.

First the receiver decodes the codewords of sender $2$. The situation is now more complicated than in case of identically distributed symbols. From the receiver's point of view the codewords of the second sender go through two different channels according to the different symbols of the codewords of the other senders. From the fact that the senders $1a$, $1b$, $2$ are synchronized the receiver knows that the first $\alpha n$ consecutive symbols of codewords go through the channel
\begin{equation}
W^{2}(y|x_{2})=\sum_{x_{1a} \in \mathcal{X}_1}\sum_{x_{1b} \in \mathcal{X}_1} \sum_{x_3 \in \mathcal{X}_3} p_{X_{1a}}(x_{1a})p_{X_{1b}}(x_{1b})p_{X_3}(x_3) W'(y|x_{1a},x_{1b},x_2,x_3),
\end{equation}
and the last $(1-\alpha)n$ consecutive symbols of codewords go through the channel
\begin{equation}
\tilde{W}^{2}(y|x_{2})=\sum_{x_{1a} \in \mathcal{X}_1}\sum_{x_{1b} \in \mathcal{X}_1} \sum_{x_3 \in \mathcal{X}_3} p_{\tilde{X}_{1a}}(x_{1a})p_{\tilde{X}_{1b}}(x_{1b})p_{X_3}(x_3) W'(y|x_{1a},x_{1b},x_2,x_3).
\end{equation}
The decoder does the following. As in Theorem \ref{egy} the $n$ tuples $(Y_{-n+1}, \dots Y_{0})$, $\dots$, $(Y_{0}, \dots Y_{n-1})$ are examined. The receiver decodes the $s$-th codeword as the $0$-th message of sender $2$ if there exists an $n$ tuple of examined output $(Y_{-n+i}, \dots Y_{i-1})$ such that the first $\alpha n$ symbols of the $s$-th codewords are jointly typical with the first $\alpha n$ symbols of $(Y_{-n+i}, \dots Y_{i-1})$ and the same is true for the last $(1-\alpha)n$ symbols according to channels $W^{2}$ and $\hat{W}^{2}$ respectively, and there are no other codewords with this property. With the shifted versions of this decoding technique the receiver also decodes the $-2,-1,1,2$-th messages of sender $2$ to ensure the decoding of the 0'th message of sender $1b$ in the last successive step. Note also that implicitly the receiver learns the delay of sender $2$ (See Remark \ref{delaydet}).

In the following successive step the receiver decodes the $-2,-1,0,1,2$-th codewords of sender $1a$ considering typicality according to the channels
\begin{equation}
W^{1a}(y,x_2|x_{1a})= \sum_{x_{1b} \in \mathcal{X}_1} \sum_{x_3 \in \mathcal{X}_3} 
p_{X_{1b}}(x_{1b})p_{X_2}(x_2)p_{X_3}(x_3) W'(y|x_{1a},x_{1b},x_2,x_3)
\end{equation}
and
\begin{equation}
\tilde{W}^{1a}(y,x_2|x_{1a})= \sum_{x_{1b} \in \mathcal{X}_1} \sum_{x_3 \in \mathcal{X}_3} 
p_{\tilde{X}_{1b}}(x_{1b})p_{\tilde{X}_2}(x_2)p_{X_3}(x_3) W'(y|x_{1a},x_{1b},x_2,x_3).
\end{equation}

In the third successive step the decoder deals with sender $3$. Note that sender $3$ is not synchronized with senders $1a$, $1b$, $2$, hence using the -2,-1,0,1,2'th codewords of the senders $2$ and $1a$ the receiver can decode (surely) just the $-1,0,1$-th codewords of sender $3$. It is also true in this case that the symbols of the codewords of sender $3$ go through two different channels:
\begin{equation}
W^{3}(y,x_{1a},x_2|x_{3})= \sum_{x_{1b} \in \mathcal{X}_1} 
p_{X_{1a}}(x_{1a})p_{X_{1b}}(x_{1b})p_{X_2}(x_2)W'(y|x_{1a},x_{1b},x_2,x_3)
\end{equation}
and
\begin{equation}
\tilde{W}^{3}(y,x_{1a},x_2|x_{3})= \sum_{x_{1b} \in \mathcal{X}_1} 
p_{\tilde{X}_{1a}}(x_{1a})p_{\tilde{X}_{1b}}(x_{1b})p_{\tilde{X}_2}(x_2)W'(y|x_{1a},x_{1b},x_2,x_3).
\end{equation}
But there is an essential difference: due to the assumption on the delays it is not known which part of the codewords goes through the channel $W^{3}$, it can be any $\alpha n$ consecutive symbols of the codewords. Here the word consecutive is understood modulo $n$. When the receiver is looking for typicality, $n^2$ joint typicality examinations are necessary according to the $n$ possible positions of the separating line of the two possible channels and the possible codeword positions.

In the final successive step the receiver decodes the $0$-th codeword of sender $1b$ considering typicality according to the channels
\begin{equation}
W^{1b}(y,x_{1a},x_2,x_3|x_{1b})= 
p_{X_{1a}}(x_{1a})p_{X_2}(x_2)p_{X_3}(x_3)W'(y|x_{1a},x_{1b},x_2,x_3)
\end{equation}
and
\begin{equation}
\tilde{W}^{1b}(y,x_{1a},x_2,x_3|x_{1b})=  
p_{\tilde{X}_{1a}}(x_{1a})p_{\tilde{X}_2}(x_2)p_{\tilde{X}_3}(x_3)W'(y|x_{1a},x_{1b},x_2,x_3).
\end{equation}

Following the calculation method of Theorem \ref{egy} it can be calculated that the rate tuple $\alpha(R_{1a},R_{1b}, R_2,R_3)+(1-\alpha)(\tilde{R}_{1a},\tilde{R}_{1b}, \tilde{R}_2,\tilde{R}_3)$ is achievable with this method. Note that a genie added version of the model is necessary to a fully rigorous error estimation, as in \cite{Urbanke}. It is also crucial that the number of joint typicality examinations is polynomial in $n$. One part of the complete calculation can be found in Appendix C.
$\blacksquare$

\section{Summary}
This paper provides a general converse for the asynchronous multiple access channels which depends on the distribution of the delays. Interesting examples of capacity regions between the simple union and its convex closure are given. These previously unknown capacity regions are: when the set of possible delays are restricted to the even numbers, and when 2 out of 3 senders are synchronized but the third is not. These examples show, that the theory of asynchronous systems is far from complete. For further results we refer to \cite{mi2}.

\section{Appendix A}

The coding theorem for totally asynchronous MAC (with two senders) was first stated in \cite{Poltyrev}, for the case when receiver knows the delays. The theorem was stated for maximal error but the converse actually proved even for average error. While the paper \cite{Poltyrev} is hard to read, with the help of reviewers of a previous version of this paper we have checked that the converse proof is correct, up to a minor gap pointed out after eq. (\ref{markov-lanc}) which is first filled here. The achievability proof in \cite{Poltyrev} is not addressed here since more accessible proofs have been published since then. In \cite{Hui-Humblet} the same capacity region was claimed to be achievable also when the receiver was uninformed. However, in the delay-detection part of the proof in Appendix 1 of \cite{Hui-Humblet} there is a gap, in eq. (12b) an independence is assumed that need not hold when the examined $n$-block of channel input symbols consists of two parts, a codeword part and a sync sequence part. The other part of the proof addresses decoding the sent codewords when the delay is already known. This part is correct, giving rise to a valid achievability proof in the case of informed receiver. Another such proof was given in \cite{Urbanke} via rate splitting and successive decoding, for any number of senders. Achievability in the uninformed receiver case has not been revisited until recently. The mentioned error in \cite{Hui-Humblet} was corrected in \cite{elgamal}. Our approach to delay detection differs from that of \cite{Hui-Humblet} and \cite{elgamal} not so much in using a sync sequence but rather in our relying on a technique from \cite{Gray} to bound the probability of delay detection error.

\section{Appendix B - Proof of Theorem \ref{main-converse-prop-alt}}

Let $S$ $\subset$ $[K]$. We will derive a bound for $R({S})$.  As in Section 3, take a window of the receiver consisting of $N+1$ n-length blocks $\mathbf{Y^{N+1}}=\{Y_0,Y_1, \dots, Y_{n(N+1)-1} \}$ and the codewords having index between $1$ and $N$ from all senders (they are fully covered by this window). Recall that $\mathbf D$ denotes the delay vector and $\mathbf X_{B,i+D_B}$ denotes the random vector  with components $X_{l,i+D_{l}}$, $l\in B$ where $B$ $\subset$ $[K]$. Denote by $\vX^{sw}$ the $2K$ input codewords which overlap with the beginning and end of $\mathbf Y^{N+1}$. Then
\begin{align}
 N&nR(S) =\\
=&\hH(\mathbf M^{N}_{S})\\
 =&\I(\mathbf M^{N}_{S} \wedge \hat{\mathbf{M}}^{N}_{S}) + \hH(\mathbf M^{N}_{S}|\hat{\mathbf{M}}^{N}_{S})\\
 \leq& \I(\mathbf M_{S}^{N} \wedge \hat{\mathbf{M}}_S^N)+Nn\eps_n \\
 \leq& \I(\mathbf X_S^N \wedge \mathbf Y^{N+1}, \vX^{sw}, \VD) +Nn\eps_n 
\end{align}
\begin{align} 
 =&\hH(\mathbf X_{S}^{N}|\VD)-\hH(\mathbf X_{S}^{N}|\mathbf Y^{N+1},\VD)+\hH(\mathbf X_{S}^{N}|\mathbf Y^{N+1},\VD)-\hH(\mathbf X_{S}^{N}|\vX^{sw},\mathbf Y^{N+1},\VD)+Nn\eps_n\\
  \leq&\hH(\mathbf X_{S}^{N}|\mathbf X^{N}_{S^c},\VD)-\hH(\mathbf X_{S}^{N}|\mathbf Y^{N+1},\mathbf X^{N}_{S^c},\VD)+\I(\vX^{sw}\wedge \vX_S^{N}|\vY^{N+1},\VD)+Nn\eps_n \\
 =& \I(\mathbf X_{S}^{N} \wedge \mathbf Y^{N+1} |\mathbf X^{N}_{S^c},\VD)+Kn\log|\iX|+Nn\eps_n \\ 
 =& \hH(\mathbf Y^{N+1}|\mathbf X^{N}_{S^c},\VD) - \hH(\mathbf Y^{N+1}|\mathbf X_{S}^{N},\mathbf X^{N}_{S^c},\VD)+Kn\log|\iX|+Nn\eps_n \\
 =&\hH(\mathbf Y^{N+1}|\mathbf X^{N}_{S^c},\VD)+Kn\log|\iX|+Nn\eps_n -\sum_{j=0}^{N} \sum_{i=0}^{n-1} \hH(Y_{nj+i}|\mathbf{Y}_1^{nj+i-1}, \mathbf X_S^N,\mathbf X^{N}_{S^c},\mathbf D)\\
 \leq& \hH(\mathbf Y^{N+1}|\mathbf X^{N}_{S^c},\VD) - \sum_{j=1}^{N-1} \sum_{i=0}^{n-1} \hH(Y_{nj+i}|\mathbf X_{[K],nj+i+D_{[K]}},\VD) +Kn\log|\iX|+Nn\eps_n. \label{Alt-converse-interrupt-1}
\end{align}

Now introduce the a random variable $\tilde Y_i$ linked to the random variables $X_{1,i\oplus D_1}$,$X_{2,i\oplus D_2}$,\dots,$X_{K,i\oplus D_K}$ by the channel $W$ for all $i\in \{0,1,\dots,n-1\}$. Then (\ref{Alt-converse-interrupt-1}) is continued as

\begin{align}
 \leq& \sum_{i=0}^{n-1} \Big[ (N-1) \hH(\tilde Y_{i}|\mathbf X_{S^c,i\oplus D_{S^c}},\VD)+ \sum_{i=0}^{n-1} \hH(Y_{i})  \notag\\
 &+\sum_{i=Nn}^{Nn+n-1} \hH(Y_{i})-(N-1)\hH(\tilde Y_i| \mathbf X_{[K],i\oplus D_{[K]}},\VD) \Big] +Kn\log|\iX|+Nn\eps_n\\
 \leq& (N-1)\sum_{i=0}^{n-1} \I(\mathbf X_{S,i\oplus D_S} \wedge \tilde Y_Q| \mathbf X_{S^c,i\oplus D_{S^c}}|\VD)+2n\log|\iY|+Kn\log|\iX|+Nn\eps_n. \label{qbevezetese}
\end{align}

Dividing by $Nn$ and going with $N$ to infinity give
\[R(S)\leq \I(\mathbf X_{S,Q\oplus D_S} \wedge \tilde Y_Q| \mathbf X_{S^c,Q \oplus D_{S^c}},Q,\VD) + \eps_n. \label{tobbdim-converse-vege}\]

This proves Theorem \ref{main-converse-prop-alt}.

\section{Appendix C - Some calculations to Theorem \ref{ketto}}

The coding/decoding task of sender $3$: the random codebook of the third sender consists of i.i.d symbols with distribution $p_{X_3}$. This codebook contains $2^{ \left( n \alpha I(X_3 \wedge Y |X_{1a,} X_2)+(1-\alpha)I(X_3 \wedge \tilde{Y} |\tilde{X}_{1a}, \tilde{X}_2) \right)-\delta'}$ codewords. $\alpha n$ consecutive symbols of an input codeword go through channel $W^{3}$, while $(1-\alpha)n$ consecutive symbols of the codeword go through channel $\tilde{W}^{3}$. Here 'consecutive' is understood modulo $n$. Let us denote by $T \subset \{0, \dots, n-1 \}$ those indices when $W^{3}$ was used. $T$ will be called separating pattern. Note that $|T|=\alpha n$ and $T$ contains consecutive numbers. The separating pattern depends on the relative delay $D$ between the synchronized senders $1a$, $1b$, $2$ and the unsynchronized sender $3$. The decoder sees an output flow (note that the symbols of senders $1a$, $2$ are also the part of the output). The decoder does not know where the codewords are separated, and does not know the separating pattern of the two possible channels. Hence the decoder should check the same output $n$ tuples as the decoder of Section 2 when looking for joint typicality, but when it examines an output $n$ tuple $Y^n$ the decoder should check every possible separating pattern. We say that the $s$'th codeword is typical in window $Y^n$ relative to separating pattern $T$ if parts of the codewords consisting of the $T$ coordinates of $\mathbf{X}^n (s)$ and $Y^n$ are jointly typical according to channel $W^{3}$, and the same is true for the  coordinates $T^{c}=\{0, \dots, n-1 \} \setminus T$ according to channel $\tilde{W}^{3}$. If $s$ is the only codeword which is typical in all the examined output windows relative to all separation patterns, then the decoder's estimation is $s$ for the 0'th message. Let us consider first the case when the examined output window is an output of the $r$'th codeword and when the real separating pattern is $T$. Then in this window the $r$'th codeword will be typical relative to $T$ with probability exponentially close to $1$ by classical arguments. First we show that no other codewords will be typical in this window. Let $T^{'}$ be any separation pattern (it can be $T$ too). We will estimate the probability that the $s \ne r$'th codeword will be typical in this window relative to $T^{'}$.

For any separation patter $T^{'}$ let $P_{X_3}^{T^{'}}(x^n,y^n)$ be the joint distribution on $\mathcal{X}^{|{T^{'}}|} \times \mathcal{Y}^{|{T^{'}}|}$ induced by the $|{T^{'}}|$-th power of $p_{X_3}$ and by the memoryless channel $W^{3}$. Let $q_{X_3}^{{T^{'}}}$ be the marginal of $P_{X_3}^{{T^{'}}}$ on $\mathcal{Y}^{|{T^{'}}|}$. Similarly let $\tilde{P}_{X_3}^{T^{'}}(x^n,y^n)$ be the joint distribution on $\mathcal{X}^{|{T^{'}}|} \times \mathcal{Y}^{|{T^{'}}|}$ induced by the $|{T^{'}}|$-th power of $p_{X_3}$ and by the memoryless channel $\tilde{W}^{3}$. Let $\tilde{q}_{X_3}^{{T^{'}}}$ be the marginal of $\tilde{P}_{X_3}^{{T^{'}}}$ on $\mathcal{Y}^{|{T^{'}}|}$. Furthermore, if $\mathbf{x}$ is an $n$-length sequence, then $\mathbf{x}^{T^{'}}$ will denote the vector of length $|T^{'}|$ consisting of those coordinates of $\mathbf{x}$ which are in $|T^{'}|$.
\begin{align}
&  \Prob_{ cond} \left\{
\left(   X_1 (s), \dots, X_n (s) ,  Y_{1},\dots, \dots, Y_{n} \right) \in S_{n}^{\delta} (T^{'}) \right\} \\
&=  \sum_{(\mathbf{x}^n (s), \mathbf{y}^n) \in S_{n}^{\delta} (T^{'})} p_{X_3}^{n}(\mathbf{x}^n (s)) \Prob_{cond} \left\{ (Y_{1}, \dots, Y_{n})=\mathbf{y}^n |(X_1 (s), \dots, X_n (s))=\mathbf{x}^n (s) \right\} \\
&=  \sum_{(\mathbf{x}^n (s), \mathbf{y}^n) \in S_{n}^{\delta} (T^{'})} p_{X_3}^{n}(\mathbf{x}^n (s)) \frac{q_{X_3}^{T^{'}}(\mathbf{y}^{T^{'}})\tilde{q}_{X_3}^{T^{'c}}(\mathbf{y}^{T^{'c}})}{q_{X_3}^{T^{'}}(\mathbf{y}^{T^{'}})\tilde{q}_{X_3}^{T^{'c}}(\mathbf{y}^{T^{'c}})}  \notag\\
&\cdot \Prob_{ cond} \left\{ (Y_{1}, \dots, Y_{n})=\mathbf{y}^n |(X_1 (s), \dots, X_n (s))=\mathbf{x}^n (s) \right\} \\
&\le \sum_{(\mathbf{x}^n (s), \mathbf{y}^n) \in S_{n}^{\delta}(T^{'})} 2^{-n \alpha \left(I(X_3 \wedge Y |X_{1a}, X_2)-\delta\right)} 2^{-n(1-\alpha)\left(I(X_3 \wedge \tilde{Y} |\tilde{X}_{1a}, \tilde{X}_2)-\delta \right)} \notag\\
&\cdot \frac{P_{X_3}^{T^{'}}(\mathbf{x}^{T^{'}} (s),\mathbf{y}^{T^{'}})\tilde{P}_{X_3}^{T^{'c}}(\mathbf{x}^{T^{'c}} (s),\mathbf{y}^{T^{'c}})}{q_{X_3}^{T^{'}}(\mathbf{y}^{T^{'}})\tilde{q}_{X_3}^{T^{'c}}(\mathbf{y}^{T^{'c}})} \cdot \notag\\
&\cdot \Prob_{ cond} \left\{ (Y_{1}, \dots, Y_{n})=\mathbf{y}^n |(X_1 (s), \dots, X_n (s))=\mathbf{x}^n (s) \right\} \\
&\le 2^{-n \alpha \left(I(X_3 \wedge Y |X_{1a}, X_2)-\delta\right)} 2^{-n(1-\alpha)\left(I(X_3 \wedge \tilde{Y} |\tilde{X}_{1a},\tilde{X}_{2})-\delta \right)}   \notag\\
&\cdot \sum_{(\mathbf{x}^n (s), \mathbf{y}^n) \in S_{n}^{\delta}(T^{'})} P_{X_3}^{T^{'}}(\mathbf{x}^{T^{'}} (s) | \mathbf{y}^{T^{'}})\tilde{P}_{X_3}^{T^{'c}}(\mathbf{x}^{T^{'c}} (s)|\mathbf{y}^{T^{'c}}) \notag\\
&\cdot \Prob_{ cond} \left\{ (Y_{1}, \dots, Y_{n})=\mathbf{y}^n |(X_1 (s), \dots, X_n (s))=\mathbf{x}^n (s) \right\}.
\end{align}

Note that $(X_1 (s), \dots, X_n (s))$ is independent from the output window ($s \ne r$), hence 1 is an upper bound of the inner sum. Note also that if one should expand $\Prob_{ cond} \left\{ (Y_{1}, \dots, Y_{n})=\mathbf{y}^n |(X_1 (s), \dots, X_n (s))=\mathbf{x}^n (s) \right\}$  then the real separation pattern $T$ should be considered. If the examined output window is related to two codewords (the $r$'th and the $l$'th codewords, where $r \ne l$) then the same argument works if $s \ne l$ and $s \ne r$. If one of them is equal to $s$ then Gray's summing technique works, the derivations can be done as in Section 4a.

\section*{Acknowledgment}
The preparation of this article would not have been possible without the
support of Dr. Imre Csisz{\'a}r. We would like to thank him for his help and advice within this subject area.

\end{document}